%% file: TOP-18-007_temp.tex
\pdfoutput=1
\documentclass[11pt,twoside,a4paper,cmspaper,final,collab]{cms-tdr}
\def\svnVersion{870467f}\def\svnDate{2023/07/18}\def\cmsCernNoTag{CERN-EP-2021-143}\def\cmsCernDate{\today}\def\cmsMessage{Published in the Journal of High Energy Physics as \href{http://dx.doi.org/10.1007/JHEP07(2023)023}{\doi{10.1007/JHEP07(2023)023}.}}
\begin{document}\cmsNoteHeader{TOP-18-007}

\newlength\cmsTabSkip\setlength{\cmsTabSkip}{1ex}

\newcommand{\ObsOne}{\ensuremath{\mathcal{O}_{1}}\xspace}
\newcommand{\ObsThree}{\ensuremath{\mathcal{O}_{3}}\xspace}
\newcommand{\ObsInd}{\ensuremath{\mathcal{O}_{i}}\xspace}
\newcommand{\AsymI}{\ensuremath{\text{A}_{\mathcal{O}_{i}}}\xspace}
\newcommand{\AsymOne}{\ensuremath{\text{A}_{\mathcal{O}_{1}}}\xspace}
\newcommand{\AsymThree}{\ensuremath{\text{A}_{\mathcal{O}_{3}}}\xspace}
\newcommand{\dtG}{\ensuremath{d_{\PQt G}}\xspace}
\newcommand{\ImdtG}{\ensuremath{\text{Im}(\dtG)}\xspace}
\newcommand{\ImdtGRatioP}{\ensuremath{(\ImdtG=2.6)/(\ImdtG=0)}\xspace}
\newcommand{\ImdtGRatioM}{\ensuremath{(\ImdtG=-2.6)/(\ImdtG=0)}\xspace}
\newcommand{\hatdt}{\ensuremath{\hat{d}_{\PQt}}\xspace}

\newcommand{\pp}{\ensuremath{\Pp\Pp}\xspace}

\newcommand{\Lag}{\ensuremath{\mathcal{L}}\xspace}
\newcommand{\gS}{\ensuremath{g_{\mathrm{S}}}\xspace}
\newcommand{\GMM}{\ensuremath{T^a}\xspace}
\newcommand{\FST}{\ensuremath{G_{\mu\nu}}\xspace}
\newcommand{\CMDM}{\ensuremath{a^{\Pg}_{\PQt}}\xspace}

\DeclareRobustCommand{\Pellp}{{\HepParticle{\Pell}{}{+}}\Xspace}
\DeclareRobustCommand{\Pellm}{{\HepParticle{\Pell}{}{-}}\Xspace}
\DeclareRobustCommand{\Pellpm}{{\HepParticle{\Pell}{}{\pm}}\Xspace}
\newcommand{\momt}{\ensuremath{p_{\PQt}}\xspace}
\newcommand{\momat}{\ensuremath{p_{\PAQt}}\xspace}
\newcommand{\momb}{\ensuremath{p_{\PQb}}\xspace}
\newcommand{\momab}{\ensuremath{p_{\PAQb}}\xspace}
\newcommand{\momlp}{\ensuremath{p_{\Pellp}}\xspace}
\newcommand{\momlm}{\ensuremath{p_{\Pellm}}\xspace}
\newcommand{\Ai}{\ensuremath{A_i}\xspace}

\newcommand{\mtop}{\ensuremath{m_{\PQt}}\xspace}
\renewcommand{\EE}{\ensuremath{\Pep\Pem}\xspace}
\renewcommand{\MM}{\ensuremath{\PGmp\PGmm}\xspace}
\newcommand{\EM}{\ensuremath{\Pepm\PGmmp}\xspace}
\newcommand{\EMinv}{\ensuremath{\Pemp\PGmpm}\xspace}
\newcommand{\Wjets}{\ensuremath{\PW\text{+jets}}\xspace}
\newcommand{\Zjets}{\ensuremath{\PZ\text{+jets}}\xspace}
\newcommand{\WW}{\ensuremath{\PW\PW}\xspace}
\newcommand{\WZ}{\ensuremath{\PW\PZ}\xspace}
\newcommand{\ZZ}{\ensuremath{\PZ\PZ}\xspace}
\newcommand{\tW}{\ensuremath{\PQt\PW}\xspace}
\newcommand{\ttV}{\ensuremath{\ttbar\text{+}\PV}\xspace}

\newcommand{\abseta}{\ensuremath{\abs{\eta}}\xspace}
\newcommand{\Irel}{\ensuremath{I_{\text{rel}}}\xspace}

\newcommand{\sigtt}{\ensuremath{\sigma_{\ttbar}}\xspace}
\newcommand{\Poi}{\ensuremath{\mathcal{P}}\xspace}
\newcommand{\BR}{\ensuremath{\mathcal{B}}\xspace}
\newcommand{\Nobs}{\ensuremath{N^{\text{obs}}}\xspace}
\newcommand{\Nobspm}{\ensuremath{\Nobs_\pm}\xspace}
\newcommand{\Nobsp}{\ensuremath{\Nobs_+}\xspace}
\newcommand{\Nobsm}{\ensuremath{\Nobs_-}\xspace}
\newcommand{\Npred}{\ensuremath{N^{\text{pred}}}\xspace}
\newcommand{\Npredpm}{\ensuremath{\Npred_\pm}\xspace}
\newcommand{\Npredp}{\ensuremath{\Npred_+}\xspace}
\newcommand{\Npredm}{\ensuremath{\Npred_-}\xspace}
\newcommand{\Ntt}{\ensuremath{N^{\ttbar}}\xspace}
\newcommand{\Nbkg}{\ensuremath{N^{\text{bkg}}}\xspace}
\newcommand{\sigeff}{\ensuremath{\varepsilon_{\text{sig}}}\xspace}
\newcommand{\fbkg}{\ensuremath{f^{\text{bkg}}}\xspace}
\newcommand{\mubkg}{\ensuremath{\mu^{\text{bkg}}}\xspace}
\newcommand{\sigbkg}{\ensuremath{\sigma^{\text{bkg}}}\xspace}

\newcommand{\hdamp}{\ensuremath{h_{\text{damp}}}\xspace}

\cmsNoteHeader{TOP-18-007}
\title{Search for CP violating top quark couplings in \texorpdfstring{\pp}{pp} collisions at \texorpdfstring{$\sqrt{s} = 13\TeV$}{sqrt(s) = 13 TeV}}

\date{\today}

\abstract{
Results are presented from a search for CP violation in top quark pair production, using proton-proton collisions at a center-of-mass energy of 13\TeV. The data used for this analysis consist of final states with two charged leptons collected by the CMS experiment, and correspond to an integrated luminosity of 35.9\fbinv. The search uses two observables, \ObsOne and \ObsThree, which are Lorentz scalars. The observable \ObsOne is constructed from the four-momenta of the charged leptons and the reconstructed top quarks, while \ObsThree consists of the four-momenta of the charged leptons and the \PQb quarks originating from the top quarks. Asymmetries in these observables are sensitive to CP violation, and their measurement is used to determine the chromoelectric dipole moment of the top quark. The results are consistent with the expectation from the standard model.
}

\hypersetup{
pdfauthor={CMS Collaboration},
pdftitle={Search for CP violating top quark couplings in pp collisions at sqrt(s) = 13 TeV},
pdfsubject={CMS},
pdfkeywords={CMS, top quark, CP violation, chromoelectric dipole moment}}

\maketitle

\section{Introduction}
The combination of charge conjugation and parity transformation (CP) is a symmetry known to be violated in the standard model (SM).
The violation of CP was initially observed in kaon decays~\cite{PhysRevLett.13.138}.
Many experiments have studied CP violation in other processes for the last three decades~\cite{CPVBelle,PhysRevLett.122.211803,PhysRevLett.110.221601}.
However, the magnitude of CP violation observed so far is not enough to accommodate the matter-antimatter asymmetry in the universe~\cite{Sakharov_1991,PhysRevLett.70.2833}, motivating searches for additional CP violation.

It has been proposed that CP violation also takes place in the production and decay of top quark pairs (\ttbar)~\cite{Gupta:2009wu,Hayreter:2015ryk}.
Several models predict CP violations, including the multi-Higgs-doublet model and the minimal supersymmetric SM; in the first model, for example, the exchanges of neutral and charged Higgs bosons can generate dipole moments from the top quark.
In the SM, only a very small amount of CP violation is expected to occur at the production vertex, as this requires at least three loops~\cite{Gupta:2009wu, Hayreter:2015ryk}.
If a nonzero chromoelectric dipole moment (CEDM) is observed at the \ttbar production vertex and the top quarks decay into Wb following the SM process, this can be a signal of CP violation in top quark pair events.

The flavor-diagonal dipole couplings between top quarks and gluons of magnetic and electric origin would be defined through a Lagrangian density
\begin{linenomath}\begin{equation}
\label{eq:cedm_L}
\Lag = \frac{\gS}{2} \PAQt \GMM \sigma^{\mu \nu} (\CMDM + i \gamma_{5} \frac{\sqrt{2}v}{\Lambda^2} \ImdtG) \PQt \FST^a,
\end{equation}\end{linenomath}
where \ImdtG is the CP-odd CEDM, and \CMDM, \gS, \GMM, and \FST represent the chromomagnetic dipole moment, the strong coupling ($\gS = \sqrt{4\pi\alpS}$),
Gell-Mann matrices, and the field strength tensor, respectively.
The scale $\Lambda$ is the scale of the new physics responsible for the CP violation of CEDM, and $v \approx 246\GeV$ is the vacuum expectation value of the Higgs field.
The discussion in Ref.~\cite{Hayreter:2015ryk} introduces various observables that probe such CP violation effects.

The studies presented in this analysis define two observables with CP-odd correlation in \ttbar events in the dilepton final state, where both W bosons decay leptonically:
\begin{linenomath}\begin{equation}
\label{eq:phy_obs}
\ObsOne = \epsilon (\momt, \momat, \momlp, \momlm) =
\begin{vmatrix}
E_{\PQt} & \momt{}_{,x} & \momt{}_{,y} & \momt{}_{,z} \\
E_{\PAQt} & \momat{}_{,x} & \momat{}_{,y} & \momat{}_{,z} \\
E_{\Pellp} & \momlp{}_{,x} & \momlp{}_{,y} & \momlp{}_{,z} \\
E_{\Pellm} & \momlm{}_{,x} & \momlm{}_{,y} & \momlm{}_{,z}
\end{vmatrix},
\end{equation}\end{linenomath}
and
\begin{linenomath}\begin{equation}
\label{eq:phy_obs1}
\ObsThree = \epsilon (p_{\PQb}, p_{\PAQb}, \momlp, \momlm) =
\begin{vmatrix}
E_{\PQb} & \momb{}_{,x} & \momb{}_{,y} & \momb{}_{,z} \\
E_{\PAQb} & \momab{}_{,x} & \momab{}_{,y} & \momab{}_{,z} \\
E_{\Pellp} & \momlp{}_{,x} & \momlp{}_{,y} & \momlp{}_{,z} \\
E_{\Pellm} & \momlm{}_{,x} & \momlm{}_{,y} & \momlm{}_{,z}
\end{vmatrix}.
\end{equation}\end{linenomath}
These observables are obtained contracting the Levi-Civita tensor $\epsilon$~\cite{Hayreter:2015ryk} with four 4-momenta, which results in the determinant of a 4$\times$4 matrix.
The components of \ObsOne and \ObsThree are the four-momenta $p$ of the charged leptons (\Pellpm), the reconstructed top quarks and antiquarks, and the \PQb quark and antiquark jets.
The elements in the 4$\times$4 matrices are therefore the energy $E$ and three-momentum components $p_x$, $p_y$, $p_z$ of these objects.
Since these observables are scalars under Lorentz transformations, they do not depend on the reference frame.

The observables \ObsOne and \ObsThree are odd under the CP transformation.
Hence, CP violation in the production of \ttbar can be tested by a measurement of the asymmetries \AsymI of the number of produced
events, $N$, with a positive and negative value in the observable \ObsInd:
\begin{linenomath}\begin{equation}
\label{eq:int_asym}
\AsymI = \frac{N (\ObsInd>0) - N (\ObsInd<0)}{N (\ObsInd>0) + N (\ObsInd<0)}.
\end{equation}\end{linenomath}
As argued in Ref.~\cite{Hayreter:2015ryk}, the asymmetries of \ObsOne and \ObsThree are the observables with highest sensitivity to CP violation, and are linearly correlated to \dtG.
While the lack of a sizable asymmetry in \ttbar production and decay would be consistent with the SM, a significant nonzero asymmetry would indicate the existence of CP violation and hence, new physics~\cite{ATWOOD20011,valencia2013cp}.
According to Ref.~\cite{Hayreter:2015ryk}, for a CEDM of Im(\dtG) = 3 the asymmetries in \ObsOne and \ObsThree can be as large as 15 and 9\%, respectively.
Asymmetries have been measured by the CMS Collaboration in the lepton+jets channel~\cite{lepjetCPV,CMS:2022voq}. 
For the observable \ObsThree in the lepton+jets channel, an additional dilution factors arises because up and down quarks from the \PW decays cannot be distinguished. 
For the dilepton channel, \ObsThree is calculated using the positively and negatively charged leptons that are well identified so that the resolution is improved.

In this paper the asymmetries of the observables \ObsOne and \ObsThree measured in the dilepton channel at 13 TeV, as well as the \ImdtG derived from those asymmetries, are presented. The tabulated results are provided in HEPData~\cite{hepdata}.

\section{The CMS Detector}
The central feature of the CMS apparatus is a superconducting solenoid of 6\unit{m} internal diameter, providing a magnetic field of 3.8\unit{T}. Within the solenoid volume are a silicon pixel and strip tracker, a lead tungstate crystal electromagnetic calorimeter (ECAL), and a brass and scintillator hadron calorimeter, each composed of a barrel and two endcap sections. Forward calorimeters extend the pseudorapidity ($\eta$) coverage provided by the barrel and endcap detectors. Muons are detected in gas-ionization chambers embedded in the steel flux-return yoke outside the solenoid.
Events of interest are selected using a two-tiered trigger system~\cite{Khachatryan:2016bia}. The first level, composed of custom hardware processors, uses information from the calorimeters and muon detectors to select events at a rate of around 100\unit{kHz} within a fixed time interval of less than 4\mus. The second level, known as the high-level trigger, consists of a farm of processors running a version of the full event reconstruction software optimized for fast processing, and reduces the event rate to around 1\unit{kHz} before data storage.
A more detailed description of the CMS detector, together with a definition of the coordinate system used and the relevant kinematic variables, can be found in Ref.~\cite{Chatrchyan:2008zzk}.

\section{Simulated events}
The \ttbar event candidates in the dilepton final state are selected and compared with simulated events to investigate the composition of processes in the data.

Simulated \ttbar events are generated at next-to-leading order (NLO) using the \POWHEG generator (v. 2)~\cite{Nason:2004rx, Frixione:2007nw, Frixione:2007vw, Alioli:2009je, Alioli:2010xd, Re:2010bp, Campbell:2014kua} at a top quark mass (\mtop) of 172.5\GeV.
The events are simulated with \PYTHIA (v. 8.219)~\cite{Sjostrand:2014zea}, referred to as \PYTHIA 8 in the following, to simulate parton showers (PS) and hadronization using the underlying-event (UE) tune CUETP8M2T4~\cite{Skands:2014pea, CMS-PAS-TOP-16-021}. The NNPDF3.0 parton distribution function (PDF) set is used to describe the proton structure.
In this analysis, \ttbar events with two oppositely charged leptons (\EE, \EM, and \MM) originating from \PW boson decays are considered as signal (``\ttbar signal'').
All other \ttbar events (\ttbar decays leading to leptonic decays of \PGt leptons, all-jet final states, or lepton+jet events) are regarded as a background (referred to as ``\ttbar other'').
The impact of not including the events where charged leptons originate from \PGt decays as ``\ttbar signal'' is small since the information contained in the lepton from a \PGt decay is diluted relative to the \PGt lepton.
In addition, CP violation could occur at the \PGt decay vertex and affect the measurement of CP violation asymmetry caused by a CEDM of the top quark.

Drell--Yan ($\qqbar \to \PZ^\ast/\PGg \to \Pellp \Pellm$, referred to as DY), \Wjets, \WW, \WZ, \ZZ, {\ttbar}+\PW, {\ttbar}+\PZ, and the \tW channel of single top quark production are regarded as the processes that can have similar decay topologies to \ttbar in the dilepton final state.
The DY process is simulated with up to four additional partons at leading-order (LO) precision using the \MGvATNLO event generator (v. 2.2.2) with the MLM jet merging prescription~\cite{Alwall2008} and the CUETP8M1 tune~\cite{Khachatryan:2015pea}.
The \ttV (where \PV refers to a \PW or \PZ) processes are simulated using \MGvATNLO, with \PYTHIA 8 for parton showering and hadronization, and the CUETP8M1 tune.
Events from \WW, \WZ, and \ZZ diboson processes are generated at LO using \PYTHIA8 and the CUETP8M1 tune.
The \tW channel is simulated with \POWHEG (v. 1) interfaced to \PYTHIA 8 using the UE tune CUETP8M2T4.
The \Wjets events are generated at NLO using \MGvATNLO and interfaced to \PYTHIA8 with the UE tune CUETP8M1.

The interactions of particles within the CMS detector are modeled using the detector simulation \GEANTfour~\cite{Agostinelli:2003hh}.

To model the effect of additional \pp interactions within the same or nearby bunch crossing (pileup), simulated minimum-bias interactions are inserted into the simulated events.
The distribution of pileup in simulation is matched to the observed data by reweighting the simulated pileup distributions, assuming a total inelastic \pp cross section of 69.2\unit{mb}~\cite{Sirunyan:2018nqx}.

The simulations are normalized to the theoretical cross sections and the integrated luminosity of the data.
The cross sections for the DY and \Wjets processes are normalized to their next-to-NLO (NNLO) prediction obtained from the \FEWZ framework (v. 3.1.b2)~\cite{PhysRevD.74.114017}.
Single top quark production is normalized to the approximate NNLO prediction~\cite{Kidonakis2014},
and diboson and \ttV processes to the NLO prediction~\cite{CAMPBELL201010,Campbell2011,bib:Maltoni:2015ena}.
The \ttbar cross section, computed at NNLO with a next-to-next-to-leading logarithmic (NNLL) accuracy for soft gluon resummation, is $832^{+20}_{-29}\,(\text{scale}) \pm 35\,(\text{PDF+}\alpS)\unit{pb}$, where the scale uncertainty comes from varying the normalization and factorization scales and the PDF+\alpS uncertainly is obtained from different PDF sets and varying \alpS~\cite{CZAKON20142930}.

\section{Event reconstruction and selection}
The data used for this analysis are from proton-proton (\pp) collisions collected at $\sqrt{s}=13\TeV$ in the CMS detector, and correspond to an integrated luminosity of 35.9\fbinv.
The event selection is optimized to select \ttbar events in which both top quarks decay to a leptonically decaying \PW boson.

The events are further selected and categorized into channels by lepton flavor (\EE, \EM, and \MM).
The selected events are required to pass a combination of single-lepton and dilepton triggers. In each event, at least one of the following triggers must be satisfied.
The single-lepton trigger requires at least one isolated electron or muon with a transverse momentum $\pt>27$ or 24\GeV and with $\abseta < 2.4$, respectively.
The same-flavor dilepton triggers require a leading-\pt lepton with $\pt>23$ or 17\GeV, and a subleading-\pt lepton with $\pt>12$ or 8\GeV.
The different-flavor dilepton triggers require either an electron with $\pt>12\GeV$ and a muon with $\pt>23\GeV$, or an electron with $\pt>23\GeV$ and a muon with $\pt>8\GeV$.

The selected events are reconstructed offline using a particle-flow (PF) algorithm~\cite{Sirunyan:2017ulk} that aims to reconstruct and identify each individual particle (PF candidate) in an event, with an optimized combination of information from the various elements of the CMS detector.

Electron candidates are reconstructed using tracking and ECAL information~\cite{Khachatryan:2015hwa} and are excluded when their ECAL clusters are located in the transition region between the barrel and endcap detectors ($1.4442<\abseta<1.5660$) because the reconstruction efficiency in this region is reduced.
A relative isolation criterion $\Irel<0.0588$ (0.0571) is applied for an electron candidate in the barrel (endcap), where the \Irel is defined as the \pt sum of all neutral hadron, charged hadron, and photon candidates
within a distance $\DR = \sqrt{\smash[b]{(\Delta\eta)^2+(\Delta\phi)^2}} = 0.3$ (where $\phi$ is the azimuthal angle in radians) from the electron candidate in $\eta$--$\phi$ space, divided by the \pt of the electron candidate, with a correction that suppresses the residual effect of pileup collisions~\cite{CMS-DP-2017-004}.
In addition, electron identification requirements are applied to reject misidentified electron candidates and candidates originating from photon conversions.
The electron candidates must have $\pt>25$ and 20\GeV for the leading and subleading candidate, respectively, within $\abseta<2.4$.

Muon candidates are reconstructed from the track information in the tracker and muon system~\cite{Sirunyan:2018fpa}. They must be in the same \pt and $\abseta$ ranges as the electron candidates above.

A relative isolation requirement of $\Irel<0.15$ is applied to muon candidates, calculated from reconstructed particles within $\DR=0.4$ of the muon in $\eta$--$\phi$ space.
In addition, muon identification requirements are used to reject muon candidates that are misidentified hadrons or come from the decay of charged pions or kaons~\cite{Sirunyan:2018fpa}.

Jets are reconstructed by clustering the PF candidates using the anti-\kt algorithm with a distance parameter $R = 0.4$~\cite{Cacciari:2008gp,Cacciari:2011ma}.
The momentum of the jet is defined as the vector sum of all particle momenta in the jet cone, and is found from simulation to be within 5 to 10\% of the true momentum over the entire \pt range of interest and detector acceptance.
Since pileup collisions contribute calorimetric energy depositions and extra tracks, tracks identified as originating from pileup vertices are discarded~\cite{Sirunyan:2017ulk}.
In addition, an offset correction is applied to jet energies in order to remove the residual energy contribution from pileup~\cite{Khachatryan:2016kdb}.
The energy scale corrections of jets are obtained from simulation and used to bring the mean measured response to that of jets at particle level.
In situ measurements of the momentum imbalance in dijet, photon+jets, \Zjets, and multijet events are used to account for any residual differences between jet energy in data and simulation~\cite{Khachatryan:2016kdb}.
Jets are required to have $\pt>30\GeV$ and $\abseta<2.4$.
If the separation \DR{} between the jet and the closest lepton is $<$0.4, the jet is discarded.

{\tolerance=800
Jets originating from the fragmentation and hadronization of \PQb quarks are identified (``\PQb tagged'') using the combined secondary vertex algorithm that uses information from track impact parameters and secondary vertices identified within a given jet.
The chosen working point provides a \PQb jet identification efficiency of approximately 63\% with a probability to misidentify light-flavor jets as \PQb jets of $\approx$1\% in \ttbar events~\cite{Sirunyan:2017ezt}.
\par}

The missing transverse momentum vector \ptvecmiss is defined as the projection on the plane perpendicular to the beam axis of the negative vector momenta sum of all reconstructed PF candidates in an event. Its magnitude is referred to as \ptmiss.

The selected events are required to have two oppositely charged isolated leptons
(\EE, \EM, or \MM) and at least two jets.
At least one of the jets is required to be \PQb tagged.
Events with additional loosely isolated electron (muon) of $\pt>20\GeV$ are vetoed. An electron is considered loosely isolated if $\Irel<0.175$ (barrel) or $0.159$ (endcaps). A muon is loosely isolated with $\Irel<0.25$.
The dilepton invariant mass must be greater than 20\GeV to suppress contributions from heavy-flavor resonance decays and low-mass DY events.
In same-flavor lepton channels, the backgrounds from \Zjets are further suppressed by rejecting events with a dilepton invariant mass within 76 and 106\GeV (\PZ boson mass window) and requiring $\ptmiss > 40\GeV$.
To estimate the remaining background contribution from \Zjets events in the same-flavor lepton channels, the events outside the \PZ mass window are normalized to the event yield in simulation within the \PZ boson mass window.
As contamination from non-DY backgrounds can still be present in the \PZ boson mass window, this contribution is subtracted with the \EM channel scaled according to the event yields in the \EE and \MM channels.
Other sources of background, such as single top quark production, diboson, \ttV, \ttbar other, misidentified leptons, and leptons within jets are estimated using simulations. In this analysis, the contribution from \Wjets events is negligible.

Since the observables in Eq.~\eqref{eq:phy_obs} contain the four-momenta of the top quark and antiquark or of the \PQb quark and antiquark, a kinematic event reconstruction~\cite{bib:TOP-12-028_paper,Sirunyan:2018ucr} is necessary.
The algorithm takes into account all combinations of jets and leptons and solves a system of equations based on the following constraints: \ptmiss is assumed to originate solely from the two neutrinos;
the invariant mass of the reconstructed \PW boson ($m_{\PW}$) is constrained to 80.4\GeV~\cite{Zyla:2020zbs};
and the invariant mass of each reconstructed top quark is constrained to 172.5\GeV.
Effects of detector resolution are taken into account through a random smearing of the measured energies and directions of the reconstructed lepton and \PQb jet candidates according to their simulated resolutions.
In addition, the assumed invariant mass of the W boson is varied according to the simulated Breit--Wigner distribution of \PW boson masses~\cite{Zyla:2020zbs}.
For a given smearing, we choose the solution of the equations for the neutrino momenta yielding the smallest invariant mass of the \ttbar system.
For each solution, a weight is calculated based on the spectrum of the true invariant mass of the lepton and \PQb jet system from top quark decays at the particle level~\cite{Sirunyan:2018ucr}.
The weights are summed over 100 smearings for each combination, and the top quark and antiquark four-momenta are calculated as the weighted average.
Among the combinations, we choose the assignment of jets and leptons that yields the maximum sum of weights. The efficiency of the kinematic reconstruction is defined as the fraction of the selected \ttbar events where a solution is found, and is about 90\% in both data and simulation. 
Using this kinematic reconstruction, the correct \PQb jet assignment is found in approximately 64\% of events.
Events without solutions for the neutrino momenta are excluded from further analysis.
Using this method, the four-momenta of the top quark and antiquark are reconstructed, as are those of the \PQb and \PAQb quark jets.
Event yields for the individual dilepton channels after event selection  are shown in Table~\ref{tab:event_yields} for the prediction of the \ttbar signal, various background processes, and the observed data.

\begin{table}[!h]
\centering\renewcommand\arraystretch{1.2}
\topcaption{Simulated event yields with their statistical uncertainties (second column) and systematic uncertainties (third column) for the three dilepton channels, after implementing event selection criteria, and normalized as described in the text. Observed selected events are also shown.\label{tab:event_yields}}
\begin{tabular}{lccc}
Process & \EE & \EM & \MM \\
\hline
\ttbar signal    & $22\,216\pm64 \pm2759$ & $104\,051\pm140\pm12\,717$ & $45\,818\pm93 \pm5732$ \\
\ttbar other     & $3\,425\pm25 \pm473$   & $16\,787\pm56  \pm2\,284$  & $7\,502\pm38 \pm1104$\\
Single top quark & $899\pm13 \pm273$      & $4\,265\pm28   \pm1\,292$  & $1\,793\pm18 \pm544$\\
DY               & $700\pm57 \pm233$      & $381\pm26      \pm117$     & $1\,627\pm95 \pm117\pm543$ \\
\ttV             & $72\pm2 \pm22$         & $302\pm4       \pm89$      & $ 144\pm3 \pm43$ \\
Diboson          & $37\pm4 \pm12$         & $100\pm7       \pm31$      & $ 70\pm6 \pm24$ \\[\cmsTabSkip]
Total prediction & $27\,350\pm90 \pm3773$ & $125\,878\pm155\pm16\,528$ & $56\,954\pm140 \pm7990$ \\[\cmsTabSkip]
Data            & $26\,961$               & $126\,549$                 & $55\,993$ \\
\end{tabular}
\end{table}

\begin{figure}[!p]
\centering
\includegraphics[width=0.45\textwidth]{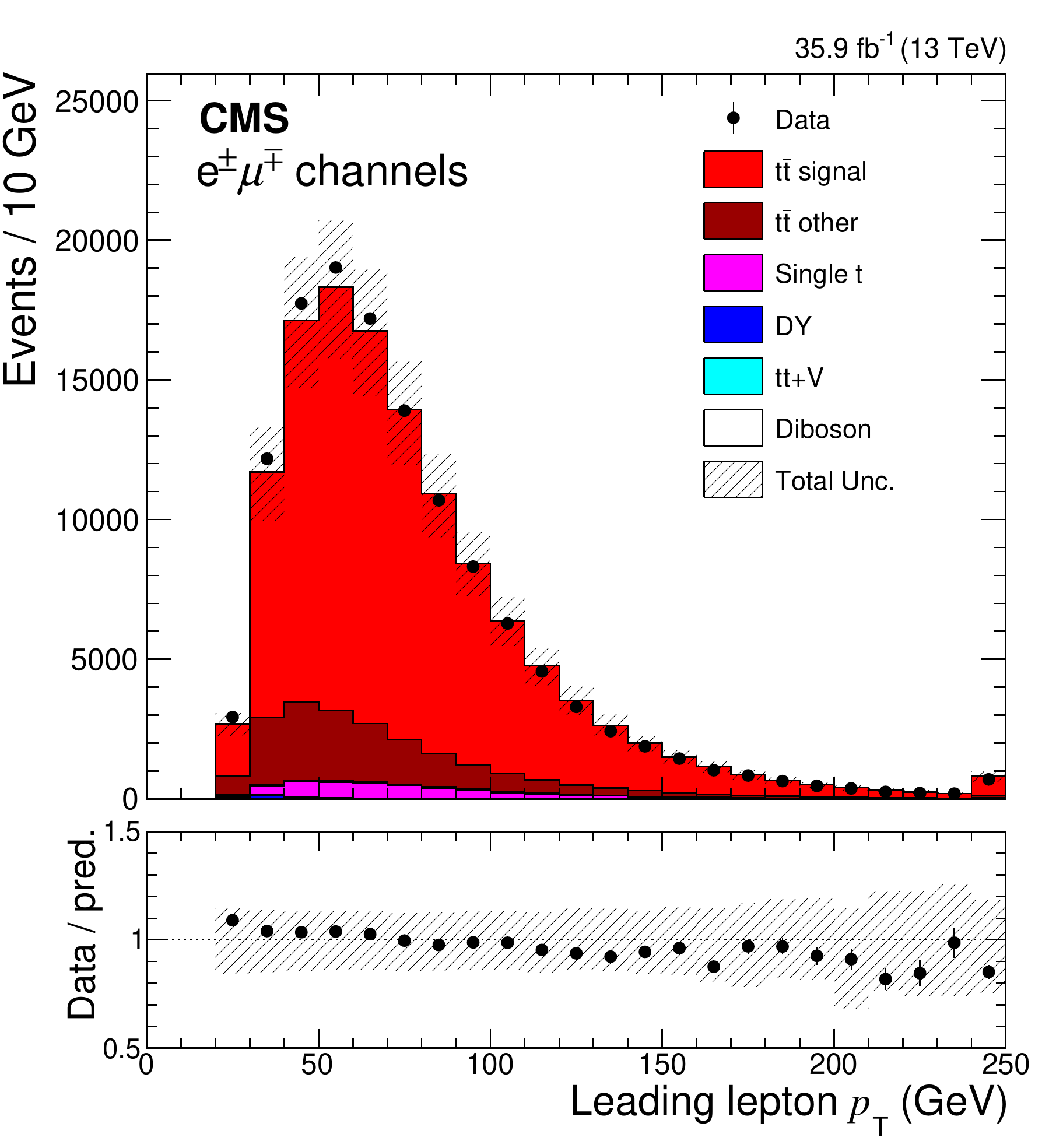}%
\hspace{0.05\textwidth}%
\includegraphics[width=0.45\textwidth]{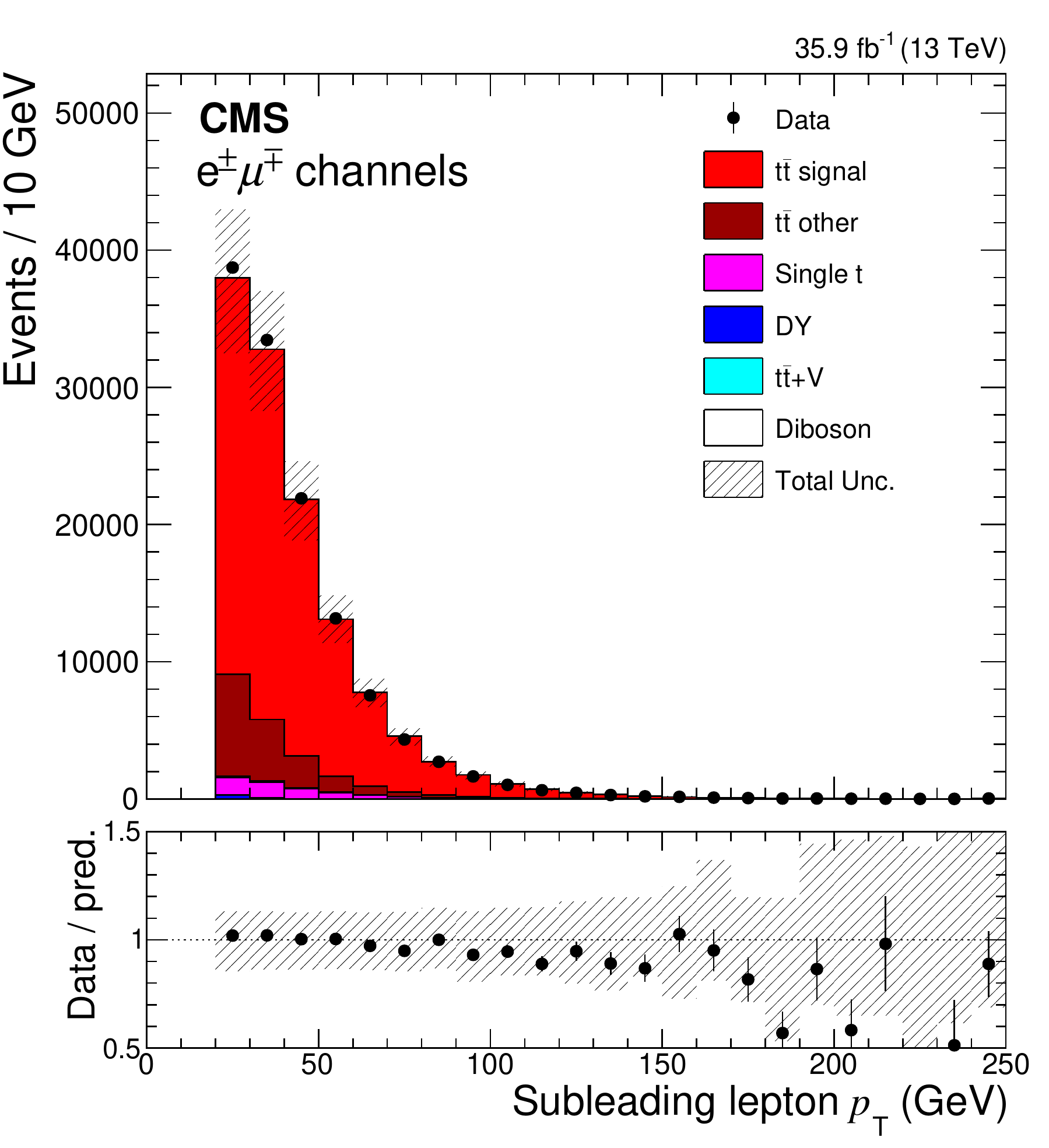} \\
\includegraphics[width=0.45\textwidth]{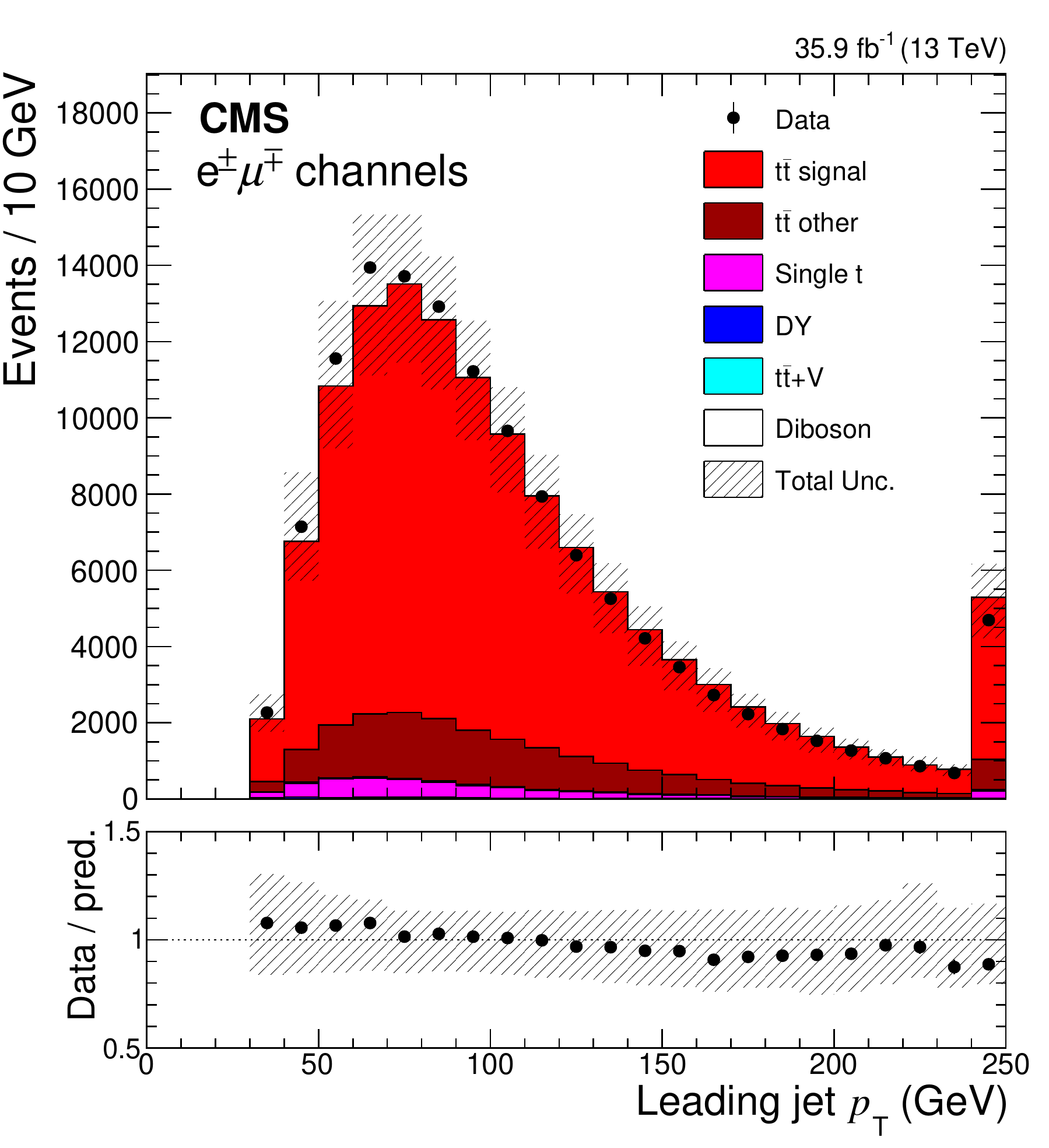}%
\hspace{0.05\textwidth}%
\includegraphics[width=0.45\textwidth]{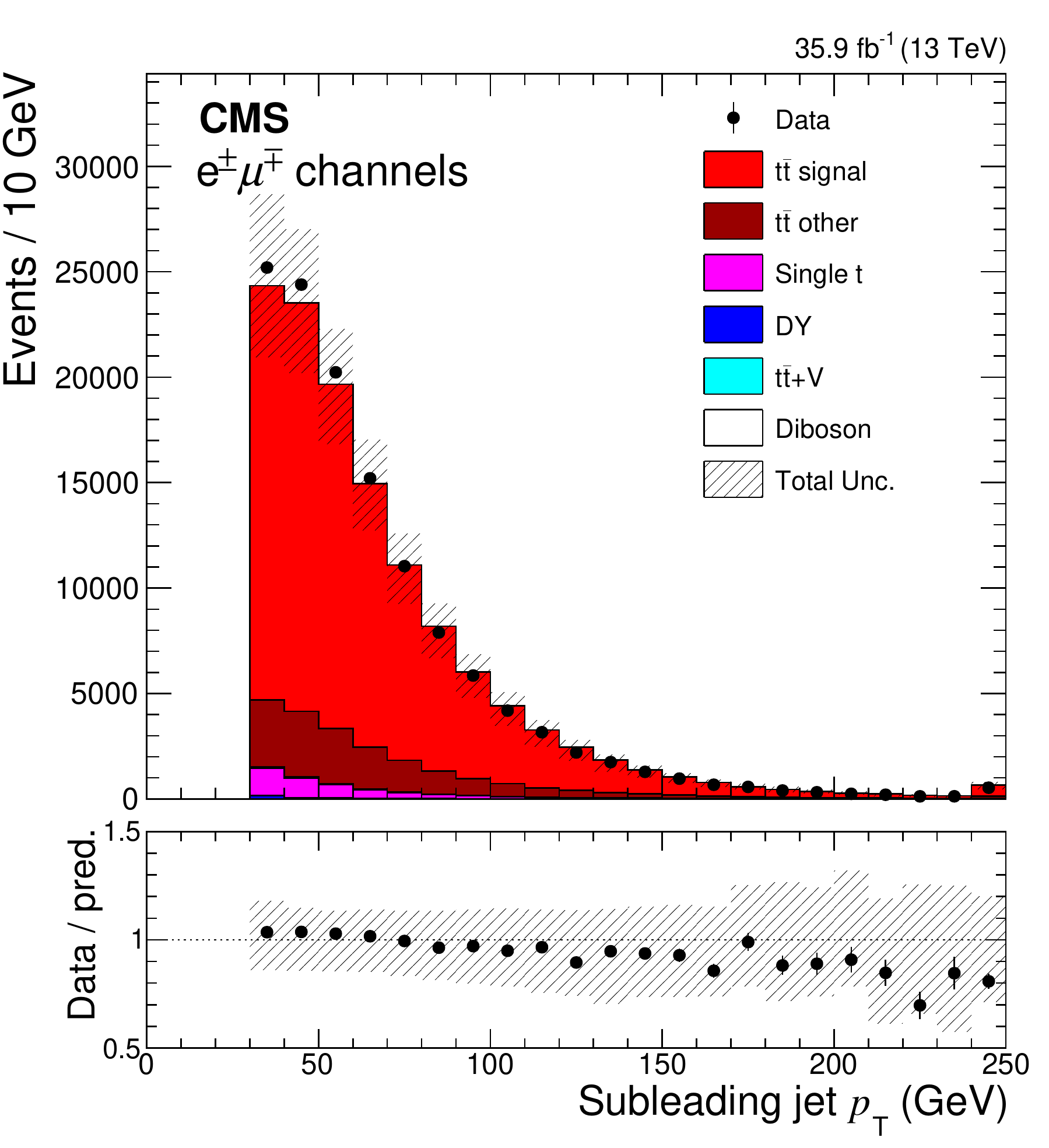}
\caption{The comparisons of the predictions and observed data in the kinematic distributions of the \pt of the leading lepton (upper left), subleading lepton (upper right), leading jet (lower left), and subleading jet (lower right) in the \EM channels.
The vertical bars on the markers of the observed data represent the statistical uncertainties.
The shaded band in the predicted distributions includes statistical and systematic uncertainties.
The last bin in each plot includes overflow events.
The ratio of the data to the predictions from simulation is presented in the lower panel of each figure.
}
\label{fig:Kine_Dist_MuEl}
\end{figure}

\begin{figure}[!p]
\centering
\includegraphics[width=0.35\textwidth]{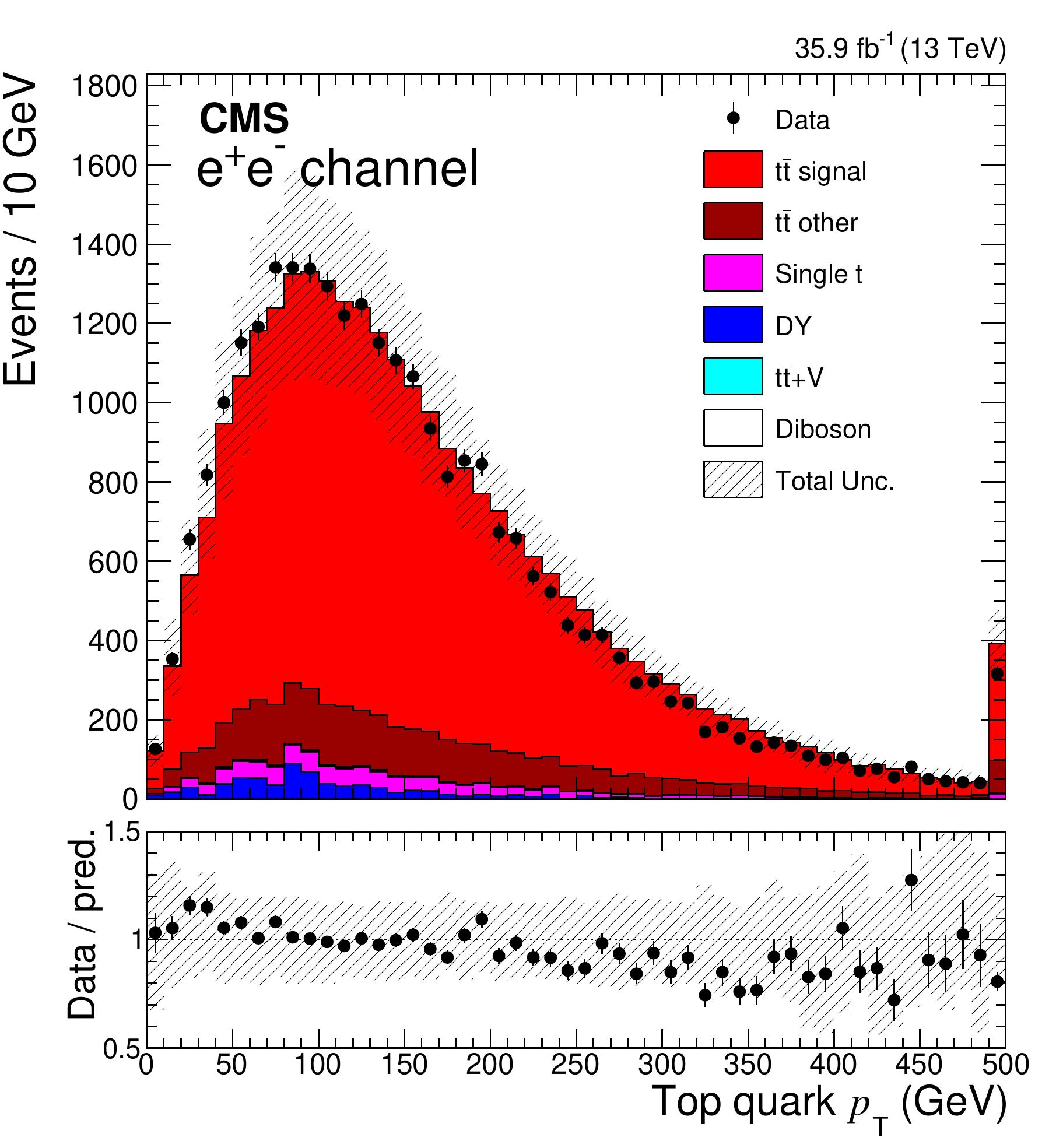}%
\hspace{0.05\textwidth}%
\includegraphics[width=0.35\textwidth]{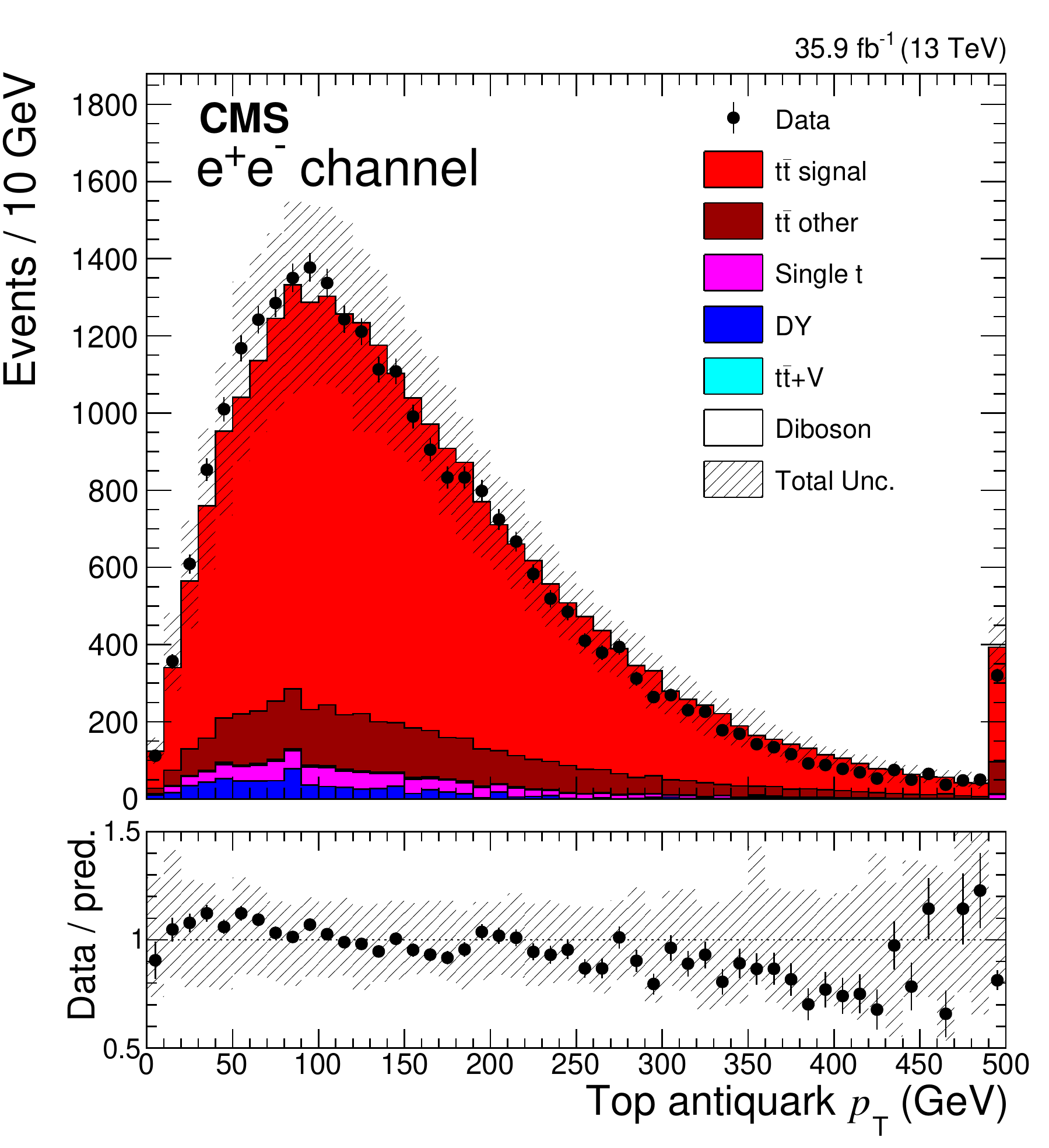} \\
\includegraphics[width=0.35\textwidth]{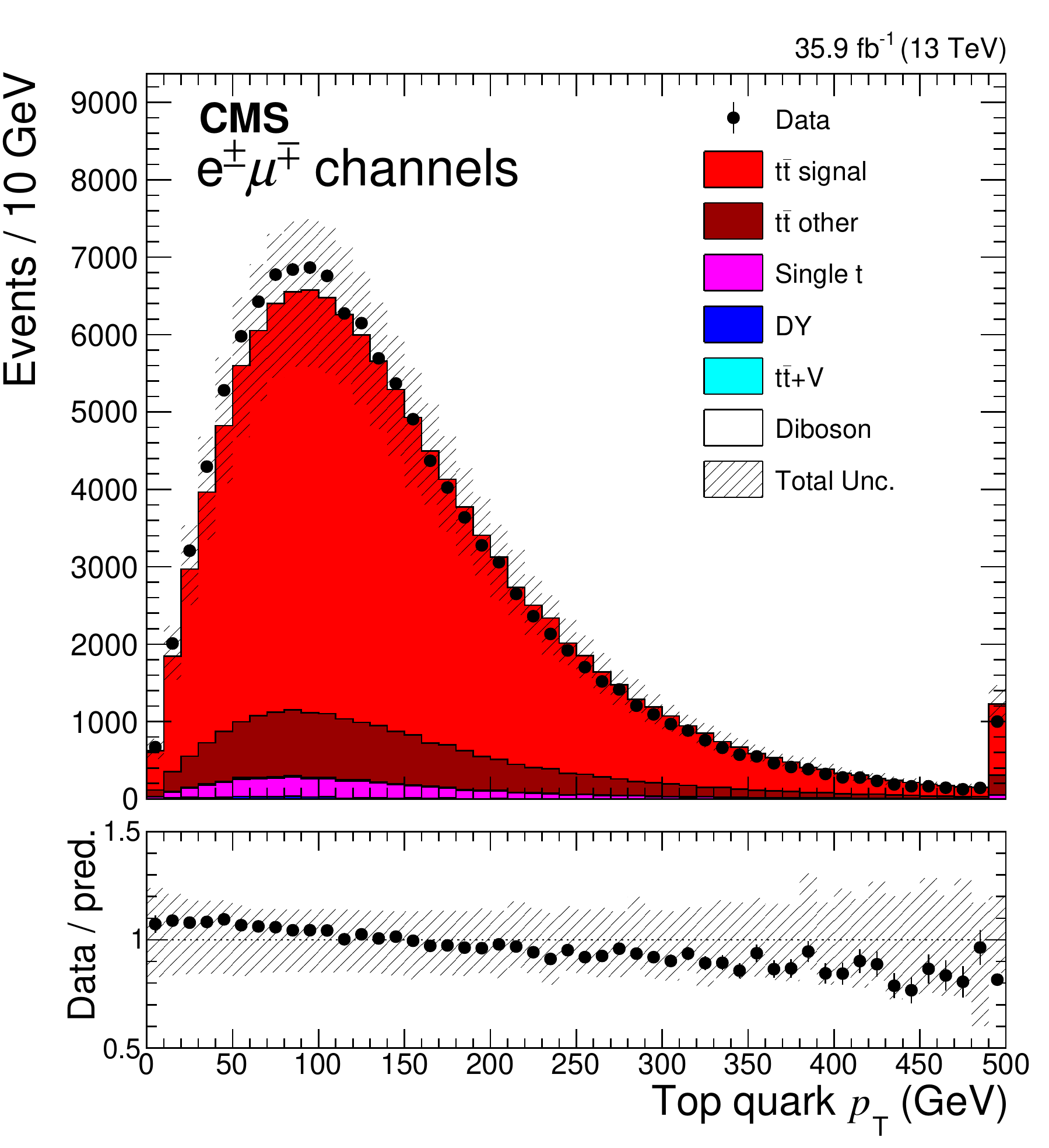}%
\hspace{0.05\textwidth}%
\includegraphics[width=0.35\textwidth]{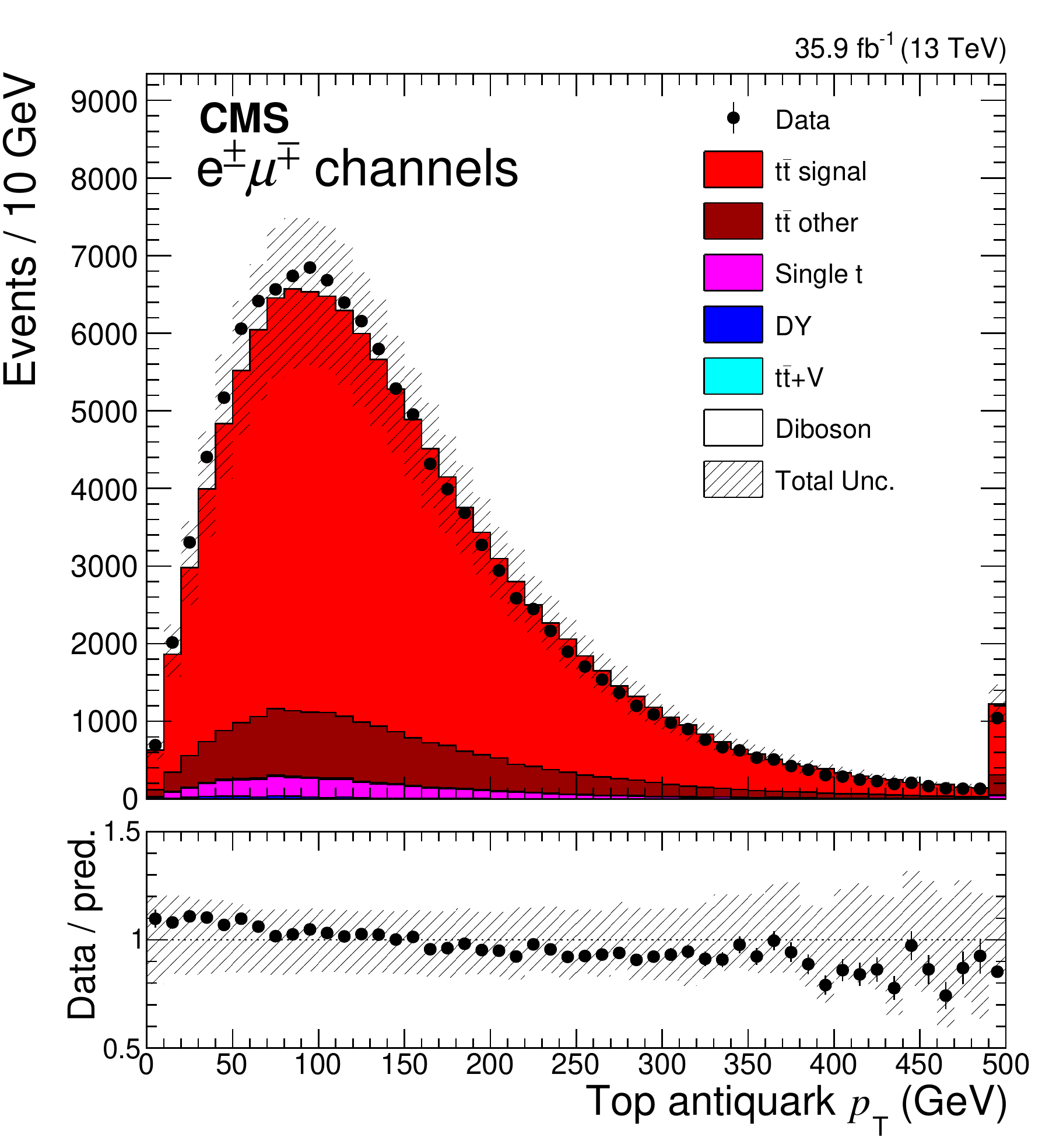} \\
\includegraphics[width=0.35\textwidth]{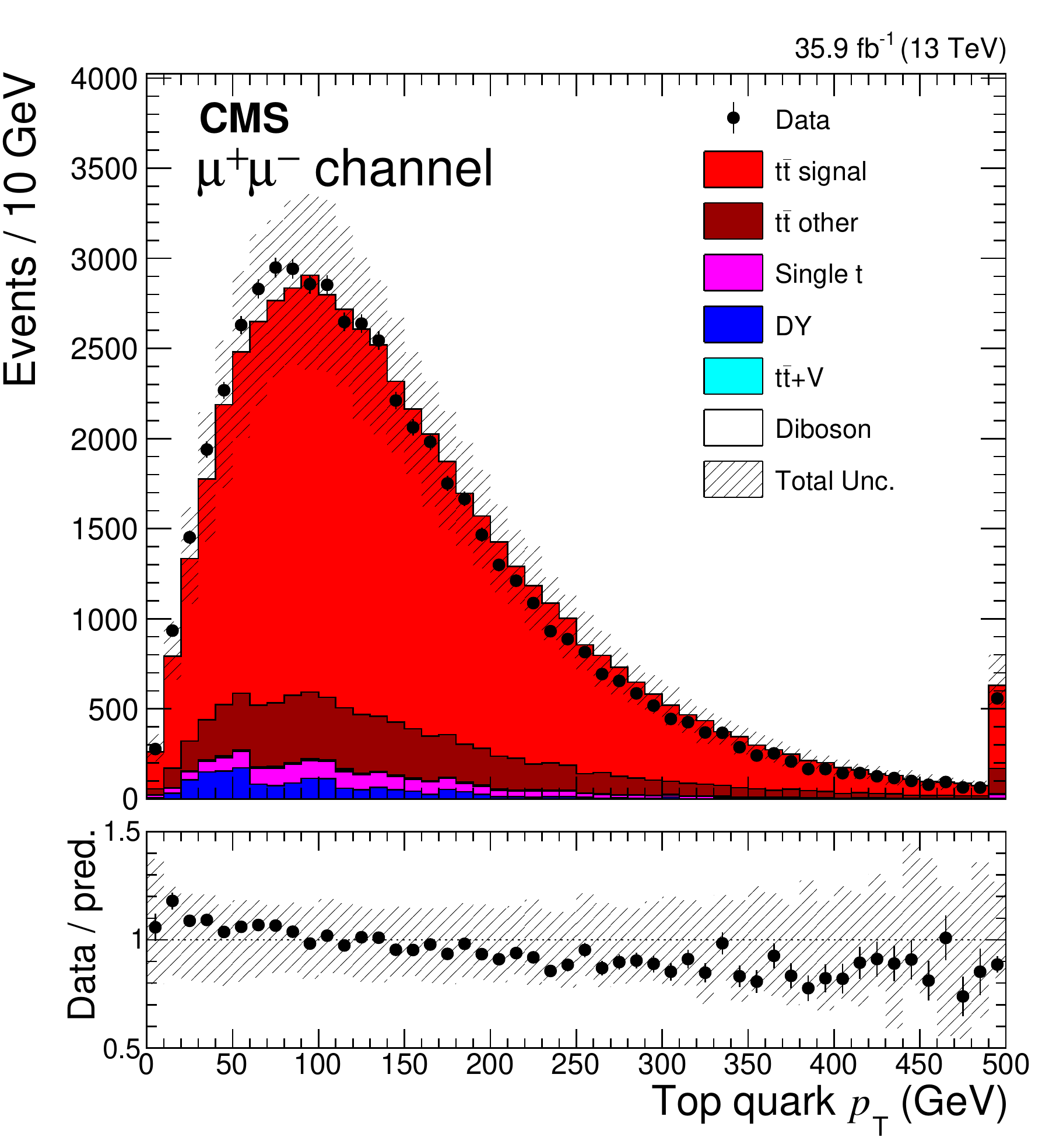}%
\hspace{0.05\textwidth}%
\includegraphics[width=0.35\textwidth]{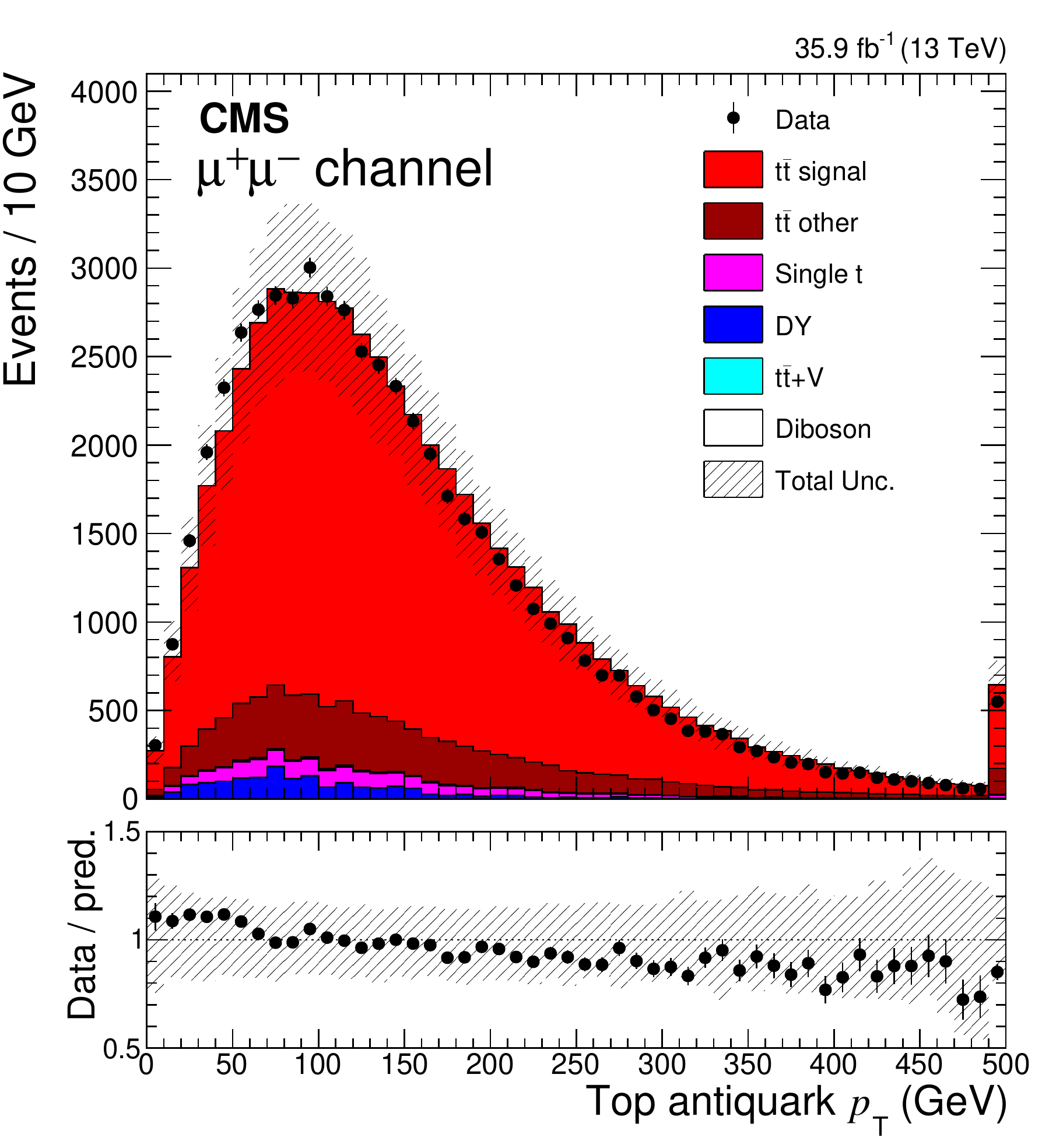}
\caption{The comparisons of the predictions and observed data in the \pt distributions in the top quark (left) and antiquark (right) in the
\EE (upper),  \EM (middle)
and \MM (lower) channels.
The vertical bars on the markers of the observed data represent the statistical uncertainties.
The shaded band in the predicted distributions includes statistical and systematic uncertainties.
The last bin in each plot includes overflow events.
The ratio of the data to the predictions from simulation is presented in the lower panel of each figure.
}
\label{fig:Kine_Dist_Toppt}
\end{figure}

\begin{figure}[!p]
\centering
\includegraphics[width=0.45\textwidth]{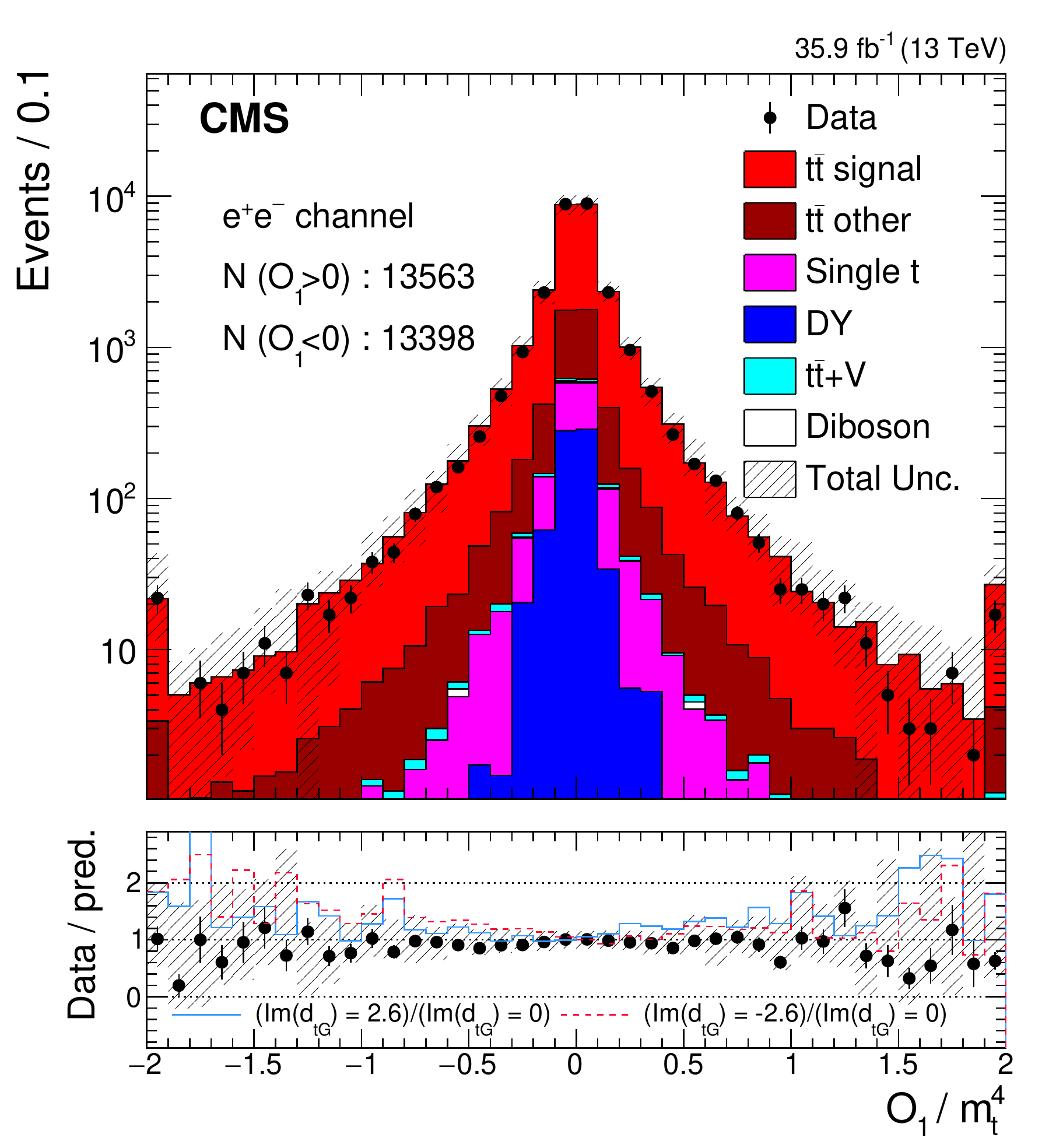}%
\hspace{0.05\textwidth}%
\includegraphics[width=0.45\textwidth]{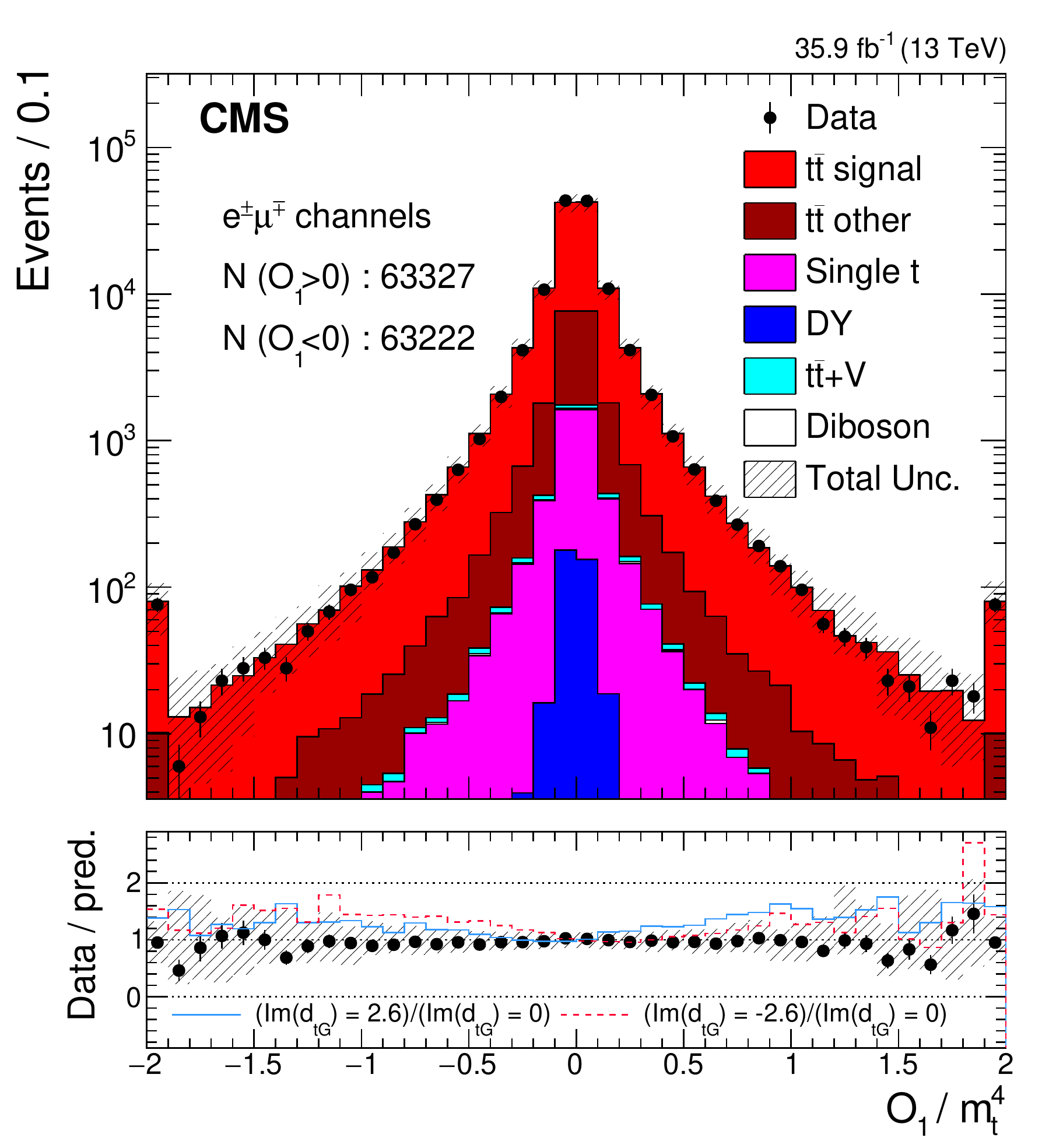} \\
\includegraphics[width=0.45\textwidth]{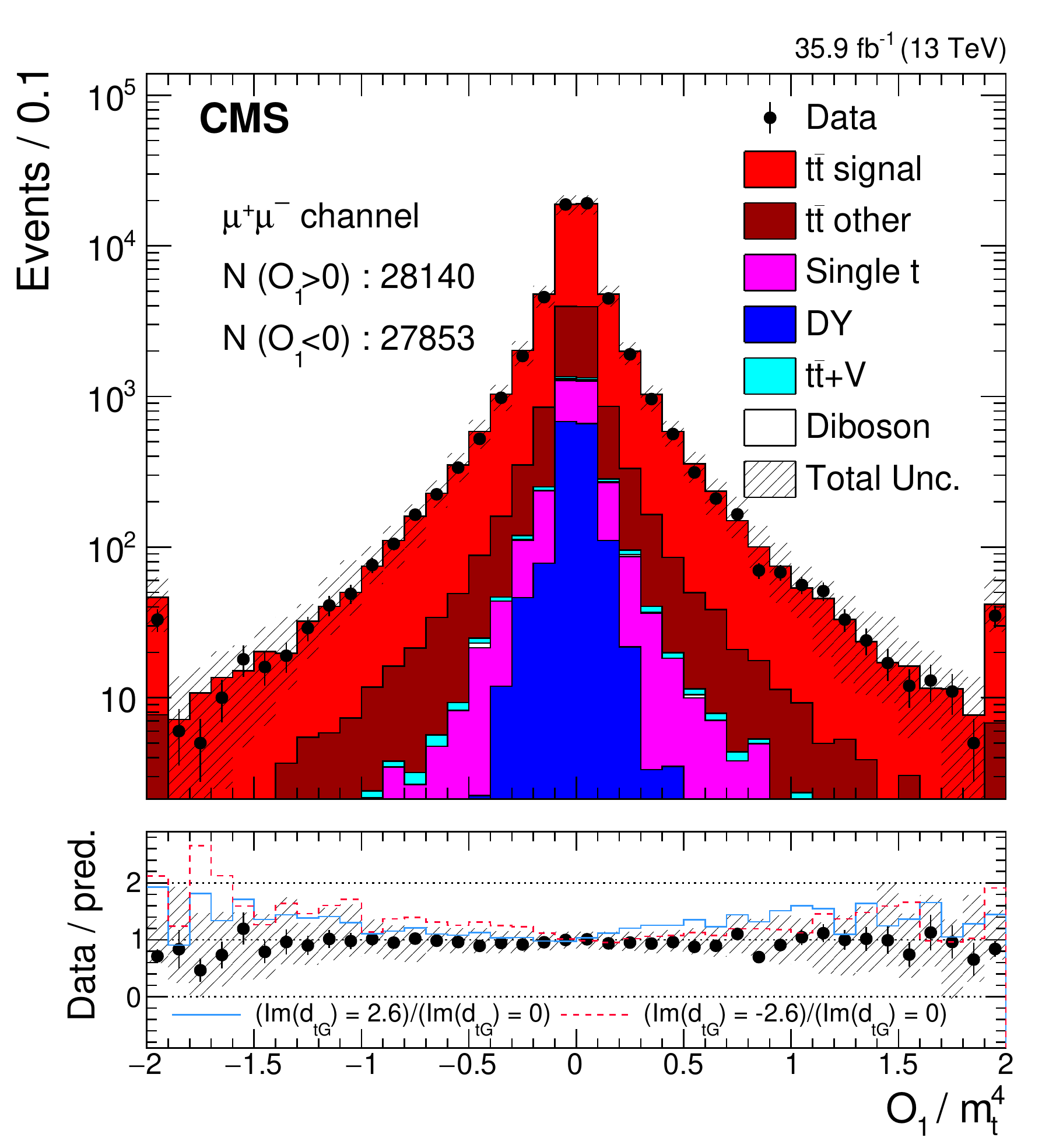}
\caption{The comparisons of the predictions and observed data in \ObsOne in the \EE (upper left), \EM (upper right), and \MM (lower) channel. 
In this figure, the values of \ObsOne are divided by the fourth power of \mtop to provide better visibility.
The vertical bars on the markers of the observed data represent the statistical uncertainties.
The shaded band in the predicted distributions includes statistical and systematic uncertainties.
The first and last bins in each plot includes underflow and overflow events, respectively.
The ratio of the data to the predictions from simulation is presented in the lower panel of each figure.
The solid blue line shows the ratio of predictions for \ImdtG = 2.6 and \ImdtG = 0,
and the dashed red line represents the ratio \ImdtG = $-2.6$ and \ImdtG = 0, using the CEDM samples.
}
\label{fig:O1_Dist}
\end{figure}

\begin{figure}[!p]
\centering
\includegraphics[width=0.45\textwidth]{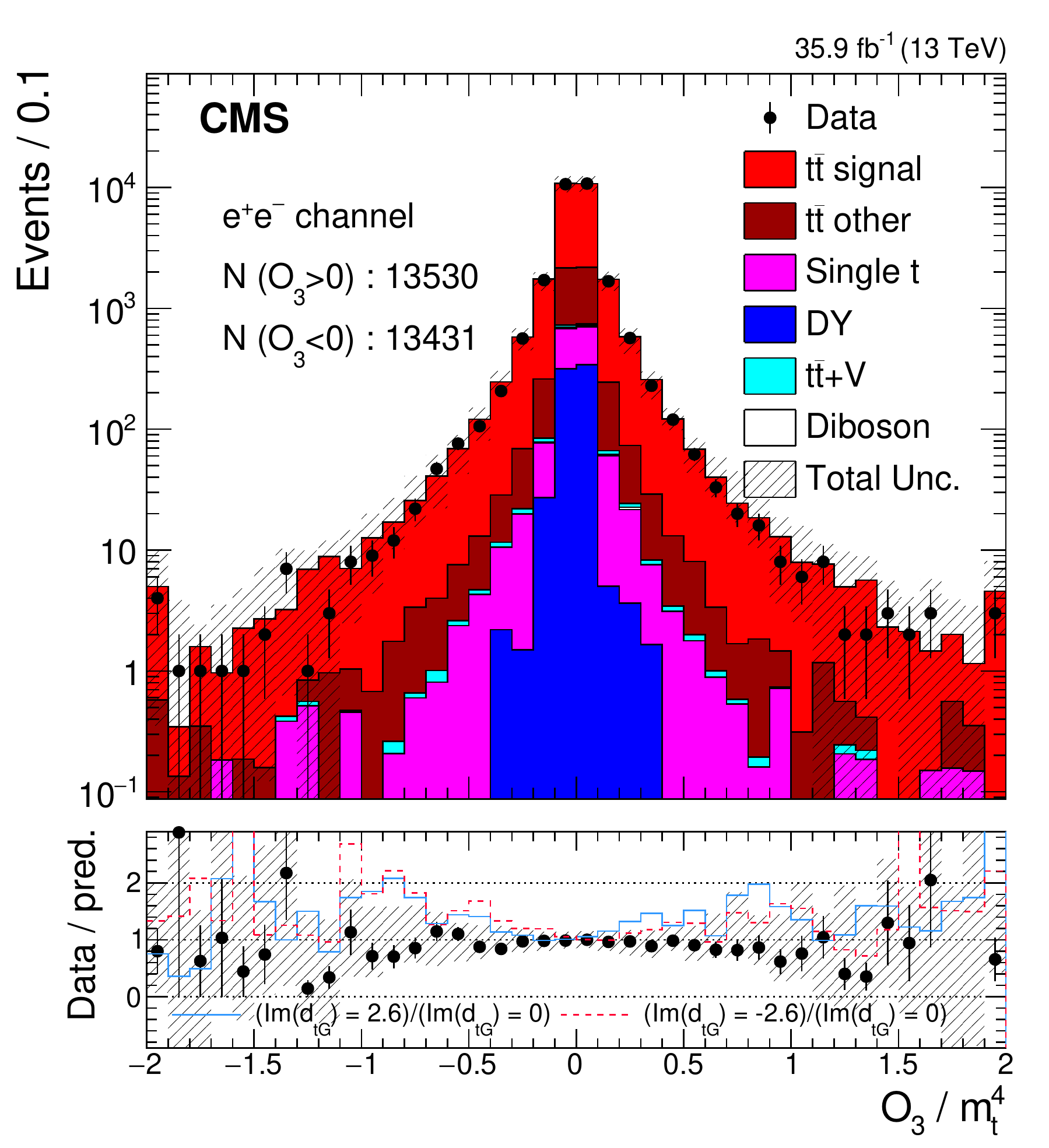}%
\hspace{0.05\textwidth}%
\includegraphics[width=0.45\textwidth]{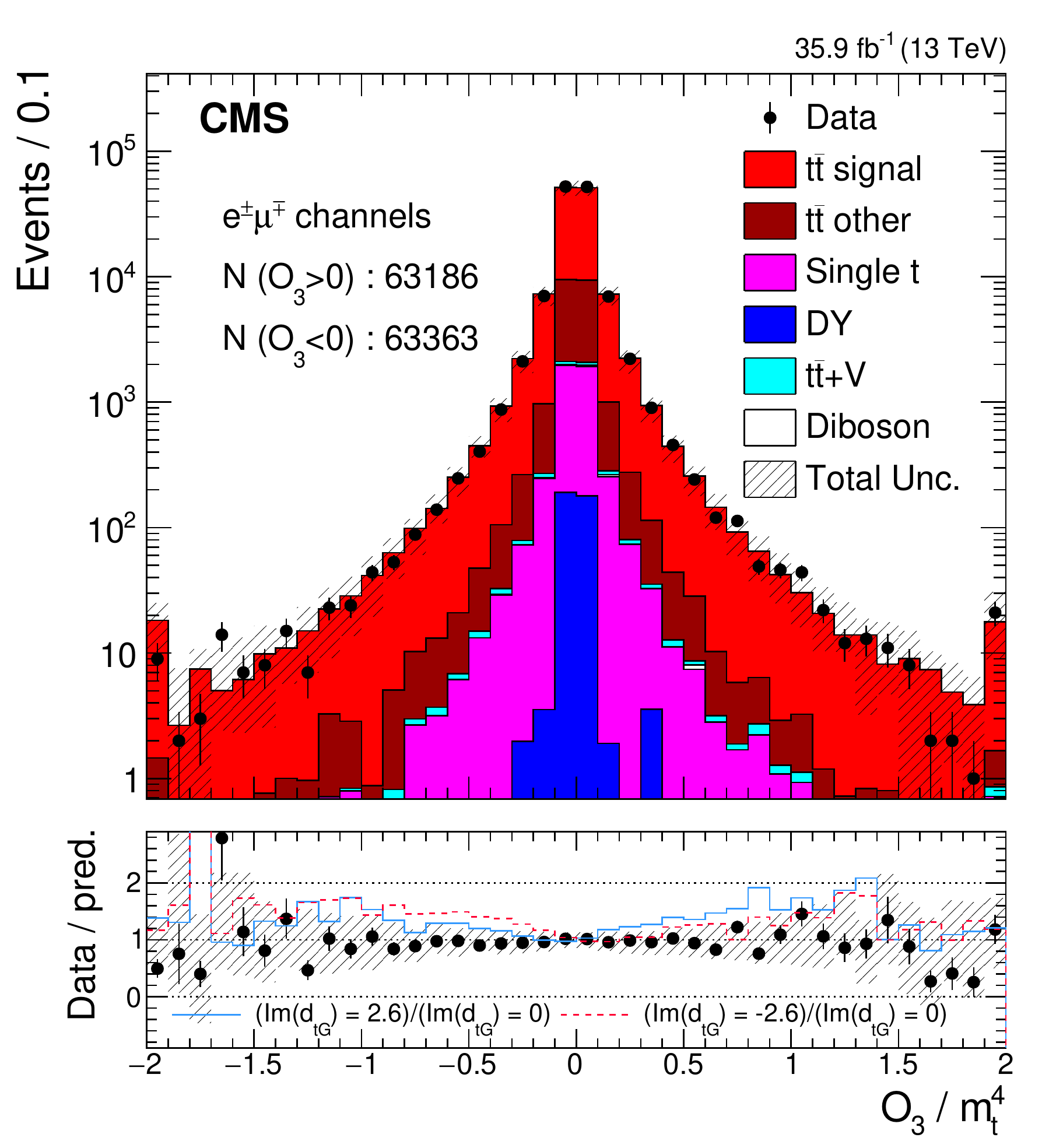} \\
\includegraphics[width=0.45\textwidth]{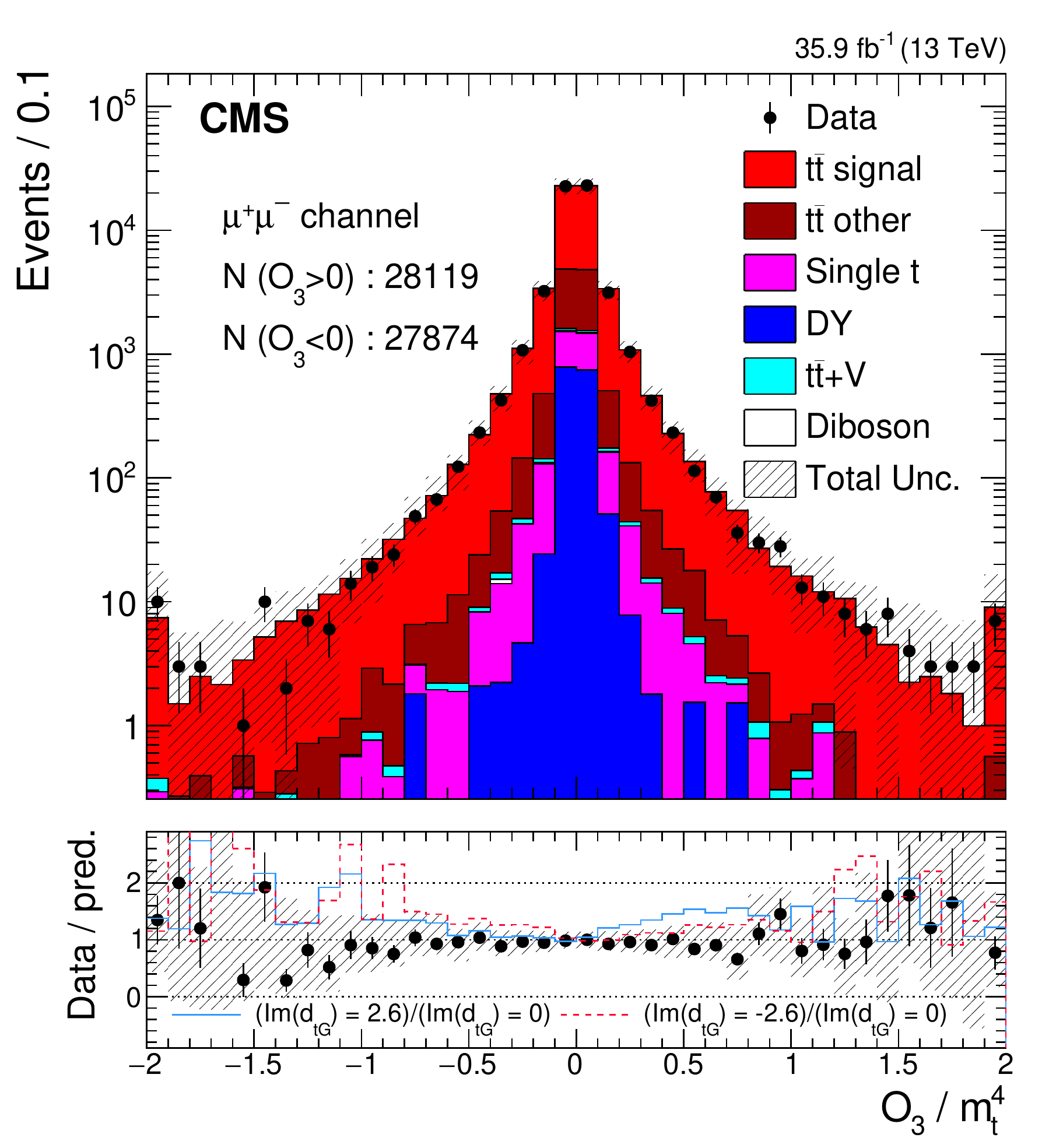}
\caption{The comparisons of the predictions and observed data in \ObsThree in the \EE (upper left), \EM (upper right), and \MM (lower) channel. 
In this figure, the values of \ObsThree are divided by the fourth power of \mtop to provide better visibility.
The vertical bars on the markers of the observed data represent the statistical uncertainties.
The shaded band in the predicted distributions includes statistical and systematic uncertainties.
The ratio of the data to the predictions from simulation is presented in the lower panel of each figure.
The first and last bins in each plot includes underflow and overflow events, respectively.
The solid blue line shows the ratio of predictions for \ImdtG = 2.6 and \ImdtG = 0,
and the dashed red line represents the ratio \ImdtG = $-2.6$ and \ImdtG = 0, using the CEDM samples.
}
\label{fig:O3_Dist}
\end{figure}

In Fig.~\ref{fig:Kine_Dist_MuEl}, distributions in the observables of the selected leptons and jets are shown for the \EM channel. The events in these distributions satisfy all event selection criteria.
Figure~\ref{fig:Kine_Dist_Toppt} shows the \pt distributions of top quark and antiquark reconstructed in the dilepton channel.
For each selected event, \ObsOne and \ObsThree are computed from the kinematic information of the reconstructed top quarks, \PQb quark jets, and leptons.
Figures~\ref{fig:O1_Dist} and~\ref{fig:O3_Dist} present the comparison of data and prediction for the observables \ObsOne and \ObsThree.
In these figures, the values of \ObsOne and \ObsThree are divided by the fourth power of \mtop to provide better visibility.
In Figs.~\ref{fig:Kine_Dist_MuEl}--\ref{fig:O3_Dist}, the shaded bands in the predicted distributions correspond
to the total uncertainty, summing in quadrature the effect of the systematic uncertainties discussed in Section~\ref{sec:Systematic uncertainties}, and the statistical uncertainties in each bin.

\section{Results}
\subsection{Extraction of asymmetries}
The distributions in Figs.~\ref{fig:O1_Dist} and~\ref{fig:O3_Dist} are split into positive and negative regions based on the algebraic sign of the observable.
The corresponding Poisson probability density functions for the observed and predicted numbers of events is used to construct a likelihood function that essentially reflects Eq.~\eqref{eq:int_asym}.
Using a maximum likelihood fit, the asymmetry \AsymI of the observable \ObsInd and the \ttbar production cross section, \sigtt, are extracted simultaneously with their statistical uncertainties.
The maximum likelihood function is defined as
\begin{linenomath}\begin{equation}
\label{eq:likelihood_func}
\Lag( \AsymI, \sigtt)=\Poi(\Nobsp, \Npredp) \, \Poi(\Nobsm, \Npredm),
\end{equation}\end{linenomath}
where the
variables \Nobspm and \Npredpm are
are the respective numbers of observed and predicted events in the positive and negative regions of \ObsInd, and $\Poi(\Nobspm, \Npredpm)$ denotes their Poisson probability density functions.
The predicted number of events \Npredpm is assumed to be a function of the two fitted parameters, \AsymI and \sigtt:
\begin{linenomath}\begin{equation}
\label{eq:n_pred}
\Npredpm = \Ntt\frac{1 \pm \AsymI}{2} + \Nbkg \fbkg_\pm,
\end{equation}\end{linenomath}
where $\Ntt = L \BR \sigeff \sigtt$,
with $L$ the integrated luminosity, \BR the dileptonic branching fraction, and \sigeff the signal efficiency.
The first term in Eq.~\eqref{eq:n_pred} reflects the number of \ttbar signal events,
where \Ntt includes the positive and negative regions.
In the second term, \Nbkg is the total number of events for all background processes combined, while $\fbkg_+$ and $\fbkg_-$ is the fraction of the number of background events in the positive (negative) regions of observable \ObsInd.
Both \fbkg and \Nbkg are fixed based on the background contributions in Table~\ref{tab:event_yields}.
Finally, the CP asymmetry and \sigtt are extracted simultaneously by minimizing the negative log-likelihood function
\begin{linenomath}\begin{equation}
\label{eq:log_likelihood}
-\log\Lag(\AsymI, \sigtt) \propto (-\Nobsp\log\Npredp + \Npredp - \Nobsm \log \Npredm + \Npredm).
\end{equation}\end{linenomath}
The resulting asymmetries with their statistical uncertainties are shown in Table~\ref{tab:asym_stat}, and the extracted cross section is consistent with previous CMS results~\cite{Sirunyan:2018ucr}.

\begin{table}[!h]
\renewcommand{\arraystretch}{1.2}
\centering
\topcaption{Measured asymmetries of \ObsOne and \ObsThree with statistical uncertainties.\label{tab:asym_stat}}
\begin{tabular}{lccc}
& \multicolumn{3}{c}{Asymmetry and uncertainty (${\times}10^{-3}$)} \\
Observable & \EE & \EM & \MM\\
\hline
\AsymOne & $8.8\pm7.5$ & $0.6\pm3.4$ & $6.9\pm5.3$  \\
\AsymThree & $4.1\pm7.5$ & $-1.7\pm3.4$ & $6.1\pm5.3$ \\
\end{tabular}
\end{table}

The asymmetries of \ObsOne and \ObsThree are statistically correlated, and the correlation factor is determined from pseudo-experiments to be 46\%.

\subsection{Systematic uncertainties}
\label{sec:Systematic uncertainties}

Systematic uncertainties may affect the asymmetries \AsymI and the CEDM values.
The effect of each systematic uncertainty is estimated by shifting the nominal prediction by the uncertainty and repeating the asymmetry measurement.
The uncertainty is taken as the average of the absolute value of the shifts of asymmetries in the up and down directions.
This allows the reduction in the statistical fluctuations in the variations.
The total systematic uncertainty is calculated by adding the effect of the individual variations in quadrature.
In this section, we discuss each of the sources assessed to be relevant in the analysis.

The uncertainty in the integrated luminosity determination is 2.5\%~\cite{bib:CMS-PAS-LUM-17-001,CMS:lumi2016}.
The uncertainty from the modeling of the number of pileup interactions is obtained by changing the inelastic \pp cross section assumed in the simulation by $\pm$4.6\%, consistent with the cross section uncertainty presented in Ref.~\cite{Sirunyan:2018nqx}.

The efficiencies of the triggers in data are measured as the fraction of events passing alternative triggers based on a \ptmiss requirement that also satisfies the criteria of the trigger of interest~\cite{TOP-17-001,Sirunyan:2018ucr}.
As the efficiency of the \ptmiss requirement is only weakly correlated with the dilepton trigger efficiencies, the bias introduced by the \ptmiss requirement is negligible.
The efficiencies are close to unity in both data and simulation, as are the corresponding data-to-simulation scale factors.
These scale factors are changed within their uncertainties to take into account the corresponding uncertainties in the efficiency.
The scale factors for the lepton identification and isolation efficiencies are determined using a tag-and-probe method~\cite{bib:tp,bib:TOP-15-003_paper}, and the uncertainties are estimated by varying the scale factors by their uncertainties.

The calibration of the electron and muon momentum scales are varied within their uncertainties, and separately for each lepton~\cite{Sirunyan:2018fpa,Khachatryan:2015hwa}.

The uncertainty arising from \PQb tagging is estimated by varying the measured \PQb tagging scale factors within one standard deviation, depending on the \pt and $\eta$ of the \PQb jets~\cite{Sirunyan:2017ezt}.
The \PQb tagging uncertainties for heavy-flavor (\PQb and \PQc) and light-flavor (\PQu, \PQd, \PQs, and gluon) jets are calculated separately, and combined in quadrature to provide the total \PQb tagging uncertainty.

The uncertainty arising from the jet energy scale (JES) is determined by changing the individual sources of uncertainty in the JES in bins of jet \pt and $\eta$, and taking their sums in quadrature~\cite{Khachatryan:2016kdb}. These changes are then propagated to the uncertainties in \ptmiss.

An additional uncertainty in the calculation of \ptmiss is estimated by varying the energies of reconstructed particles not clustered into jets.

The uncertainty originating from the jet energy resolution (JER) is determined by changing the JER in simulation within its uncertainty within different $\eta$ regions~\cite{Khachatryan:2016kdb}.

The simulated background samples were generated with a limited number of events.
Therefore, the total number of background events is affected by statistical fluctuations in these samples.
The fluctuations in asymmetries originating from  such limitations are evaluated through nuisance parameters in the likelihood function associated with the numbers of the background events in the negative and positive regions of the observables. The likelihood function in Eq.~\eqref{eq:likelihood_func} can be modified as:
\begin{linenomath}\begin{equation}
\label{eq:likelihood_nui}
\Lag(\AsymI, \sigtt)=\Poi(\Nobsp, \Npredp) \, \Poi(\Nobsm, \Npredm) \, G(\Nbkg_-|\mubkg_-, \sigbkg_-) \, G(\Nbkg_+|\mubkg_+, \sigbkg_+),
\end{equation}\end{linenomath}
where $\mubkg_\pm$ and $\sigbkg_\pm$ are the mean and the statistical uncertainty of the expected background processes.
Gaussian constraints are imposed on the nuisance parameters.

The uncertainty in the normalization of the expected background processes other than \ttbar is estimated through scaling the background yield in simulation up and down by 30\%~\cite{bib:TOP-15-003_paper}, and then extracting the asymmetries.

The uncertainty in the PDFs used to simulate the \ttbar production is obtained using the NNPDF3.0 PDFs~\cite{Butterworth:2015oua}.
The impact of the uncertainty in the renormalization and factorization scale in the \ttbar simulation is estimated by varying the factorization and renormalization scales used during the generation of the simulated sample independently by factors of 0.5 and 2.
The extreme cases where one scale is varied up, while the other one is varied down, are not considered.

The dependence of the asymmetries and of the CEDM on the assumed value of \mtop is estimated by varying the generated \mtop in the simulation by $\pm$1\GeV with respect to the default value of 172.5\GeV.

Previous studies have shown that the \pt distribution of the top quark in data is softer than expected in the NLO simulation of \ttbar production~\cite{bib:TOP-11-013_paper,bib:TOP-12-028_paper,Khachatryan:2015fwh,bib:TOP-16-008,Sirunyan:2018ucr}.
A reweighting procedure based on the \pt spectrum of the  top quark is applied to the nominal \POWHEG prediction at NLO on an event-by-event basis so as to match the simulated spectrum of the top quark \pt to data. The change in the result is taken as the systematic uncertainty.

The uncertainty in the \PQb jet modeling has three components. The fragmentation into \PQb hadrons is varied in simulation within the uncertainties of the Bowler--Lund fragmentation function tuned to ALEPH~\cite{APLHE} and DELPHI~\cite{DELPHI} data.
In addition, the difference between the Bowler--Lund \cite{Bowler:1981sb} and the Peterson \cite{PhysRevD.27.105} fragmentation functions is included in the uncertainty.
Lastly, the uncertainty from the semileptonic \PQb hadron branching fraction is obtained by varying it by $-0.45$ and $+0.77\%$, which is the range of the measurements from $\PBz/\PBp$ decays and their uncertainties \cite{Zyla:2020zbs}.

The uncertainty in the matching scale between the matrix element (ME) and the parton shower is evaluated by varying the \hdamp parameter that regulates the emissions in \POWHEG.
The nominal value of \hdamp in the simulation is $1.58\mtop$, and the modified values are $0.99\mtop$ and $2.24\mtop$, obtained from tuning the parameter using \ttbar data at $\sqrt{s}=8$ and 13\TeV~\cite{CMS-PAS-TOP-16-021}.

The default setup in \PYTHIA includes a multiple-parton interaction (MPI) scheme of the color reconnection (CR) model with ``early'' resonance decays switched off.
To estimate the uncertainty from this choice of model, the analysis is repeated using three other CR models within \PYTHIA:
the MPI-based scheme with early resonance decays switched on, a gluon-move scheme~\cite{Argyropoulos:2014zoa},
and a scheme~\cite{Christiansen:2015yqa} inspired by quantum chromodynamics (QCD).
The total uncertainty from CR modeling is estimated by taking the maximum deviation from the nominal result.

The uncertainty from the modeling of the underlying event is estimated by varying the parameters that govern the underlying event modeling of the PS tune CUETP8M2T4 in \PYTHIA8 up and down within their uncertainties~\cite{Skands:2014pea,CMS-PAS-TOP-16-021}. 

The renormalization scale for initial- and final-state gluon radiation (ISR and FSR) is varied up and down by a factor of 2 (for ISR) and $\sqrt{2}$ (for FSR) in order to estimate the uncertainty attributed to the shower model~\cite{Skands:2014pea}. 

The observables \ObsOne and \ObsThree in Eqs.~\eqref{eq:phy_obs} and~\eqref{eq:phy_obs1} are related to Levi-Civita tensors contracted with the four-momenta of \PQt, \PAQt, \PQb, and \PAQb quark jets, and two leptons.
These four-momenta and the observables are affected by the uncertainties in detector measurements.
A fluctuation in a measurement caused by detector response effects can change the sign of the observables in an event, and as a consequence dilute the asymmetries.
We have considered the charge misidentification as additional systematic source that can dilute the asymmetries.

Charge misidentification:
the tracks of charged particles at  high momenta become almost straight lines in the 3.8\unit{T} magnetic field.
Fluctuations in the measured hit positions along the track can change the sign of the sagitta and thereby change the sign of the charged track.
In such cases, the lepton charge can be misidentified, which would result in a dilution of the asymmetry.
We identify two ways of misidentifying lepton charge.
In the first case, the charge of one lepton is assumed to be correctly identified while the charge of the other lepton is misidentified.
Since the leptons would have the same charge, events would be rejected by the \ttbar signal selection requirement of oppositely charged lepton pairs.
The probability of the charge of one lepton being misidentified is calculated separately for the positive and negative region of the observables.
Based on this probability, events are vetoed and the asymmetry measurements are repeated.
In the second case, the charges of both leptons are assumed to be misidentified.
In this case, the positions of the four-momenta of \Pellp and \Pellm are swapped in the Levi-Civita tensor.
Consequently, the sign of the tensor changes because of the properties of the matrix determinant.
The measured misidentification probability is applied to measured rate is applied to the MC to obtain the probability of misidentification in both the positive and negative regions.
Assuming the worst-case scenario in which events in only one of the regions contain two leptons with misidentified charge and migrate into the other region, the number of events in the regions can be re-estimated.
The changes in asymmetries in the individual channels are taken as uncertainty on the asymmetry arising from charge misidentification.
The likelihood for the lepton charge to be mismeasured is described as a function of the \pt and $\eta$ of the lepton. For an electron (muon) with a \pt between 20 and 50 \GeV and an \abseta less than $2.4$, the charge misidentification rate is about $0.17\%$ ($0.03\%$). 
Therefore, the probability of misidentifying two central charged leptons (and thus swapping the sign of \ObsOne/\ObsThree) is $3 \times 10^{-6}$ in the dielectron channel and negligible in the $\Pe\PGm$ and dimuon channels. In the endcaps, the charge misidentification probability is up to $2\%$ for electrons, which is well described by our MC simulation, as checked using the tag-and-probe method in \PZ boson events.

The asymmetries measured in the three dilepton channels are combined using the best linear unbiased estimator method~\cite{LYONS1988110}, taking into account the correlation of the systematic uncertainties across channels.
The systematic uncertainties are assumed to be fully correlated across channels,
except for the uncertainty in the limited number of simulated background events and in the normalization of the background models.
Since these systematic sources are statistically independent, they can be taken as uncorrelated uncertainties.
The statistical uncertainties are assumed to be uncorrelated.
The contributions of systematic uncertainties and additional sources are summarized in Table~\ref{tab:syst_table_asym_O1}.

\begin{table}[!t]
\renewcommand{\arraystretch}{1.2}
\centering
\topcaption{Systematic uncertainties in the measured asymmetries of \ObsOne and \ObsThree, for the individual and combined channels.}
\label{tab:syst_table_asym_O1}
\begin{tabular}{lcccccccc}
& \multicolumn{8}{c}{Uncertainty (${\times}10^{-3}$)} \\
& \multicolumn{2}{c}{\EE} & \multicolumn{2}{c}{\EMinv} & \multicolumn{2}{c}{\MM} & \multicolumn{2}{c}{Combined} \\
Source & \AsymOne & \AsymThree & \AsymOne & \AsymThree & \AsymOne & \AsymThree & \AsymOne & \AsymThree \\
\hline
Luminosity & 0.0 & 0.0 & 0.0 & 0.0 & 0.0 & 0.0 & 0.0 & 0.0 \\
Pileup     & 0.3 & 0.0 & 0.0 & 0.0 & 0.1 & 0.8 & 0.1 & 0.2 \\
Lepton Identification & 0.1 & 0.0 & 0.0 & 0.0 & 0.1 & 0.1 & 0.0 & 0.0 \\
Lepton Isolation      & 0.0 & 0.0 & 0.0 & 0.0 & 0.0 & 0.0 & 0.0 & 0.0 \\
Trigger               & 0.0 & 0.0 & 0.0 & 0.0 & 0.0 & 0.0 & 0.0 & 0.0 \\
Electron momentum scale/smearing & 1.2 & 1.1 & 0.2 & 0.2 & \NA & \NA & 0.3 & 0.2 \\
Muon momentum scale & \NA & \NA & 0.1 & 0.2 & 2.3 & 1.0 & 0.5 & 0.3 \\
JES                                      & 1.9 & 0.6 & 0.1 & 0.2 & 2.3 & 0.7 & 0.7 & 0.3 \\
JER                                      & 2.0 & 0.7 & 0.3 & 0.2 & 1.2 & 0.3 & 0.6 & 0.2 \\
b tagging                                & 0.1 & 0.0 & 0.0 & 0.0 & 0.1 & 0.1 & 0.0 & 0.0 \\
Normalization of background models       & 0.3 & 0.0 & 0.0 & 0.0 & 0.1 & 0.1 & 0.0 & 0.0 \\
Limited simulated background sample size & 2.9 & 2.9 & 0.6 & 0.6 & 2.3 & 2.3 & 0.7 & 0.7 \\
PDF                                      & 0.1 & 0.3 & 0.2 & 0.2 & 0.2 & 0.2 & 0.1 & 0.1 \\
ME--PS matching                          & 0.8 & 1.4 & 0.3 & 0.7 & 0.8 & 1.5 & 0.4 & 0.9 \\
Color reconnection & 1.9 & 3.8 & 1.6 & 1.0 & 1.0 & 0.9 & 1.5 & 1.1 \\
Underlying event & 0.6 & 0.9 & 1.4 & 1.1 & 1.4 & 1.0 & 1.4 & 1.0 \\
ISR & 1.5 & 1.8 & 0.2 & 0.2 & 0.3 & 0.5 & 0.3 & 0.3 \\
FSR & 1.0 & 1.9 & 0.8 & 0.6 & 0.6 & 0.3 & 0.7 & 0.6 \\
Hadronization & 2.0 & 0.5 & 0.6 & 0.3 & 1.7 & 0.2 & 0.9 & 0.3 \\
\mtop         & 0.9 & 0.1 & 0.4 & 0.2 & 0.0 & 0.1 & 0.4 & 0.1 \\
top quark \pt & 0.0 & 0.0 & 0.0 & 0.0 & 0.0 & 0.0 & 0.0 & 0.0 \\
Charge misidentification & 0.8 & 0.8 & 0.4 & 0.4 & 0.1 & 0.1 & 0.3 & 0.3 \\ [\cmsTabSkip]
Total systematic uncertainty & 5.6 & 6.0 & 2.6 & 2.0 & 5.0 & 3.5 & 2.8 & 2.2 \\
\end{tabular}
\end{table}

\subsection{Extraction of CEDM}
\label{sec:Extraction of CEDM}

As indicated in Eq.~\eqref{eq:cedm_L}, the \ttbar production vertex is modified by the presence of CEDMs, leading to CP violation.
According to Ref.~\cite{Hayreter:2015ryk}, the asymmetry defined in Eq.~\eqref{eq:int_asym} is linearly proportional to the CEDM.
Dedicated
\ttbar events
with such CP modifications in Eq.~\eqref{eq:cedm_L} are generated using \MGvATNLO interfaced to \PYTHIA 8~\cite{Sjostrand:2014zea} using the UE tune CUETP8M2T4.

Seven samples are generated with different values of the dimensionless CEDM \dtG parameter (which are $-2.6$, $-1.0$, $-0.5$, $0$, $0.5$, $1.0$, $2.6$). These samples include the CMS full detector simulation.

\begin{figure}[!ht]
\centering
\includegraphics[width=0.45\textwidth]{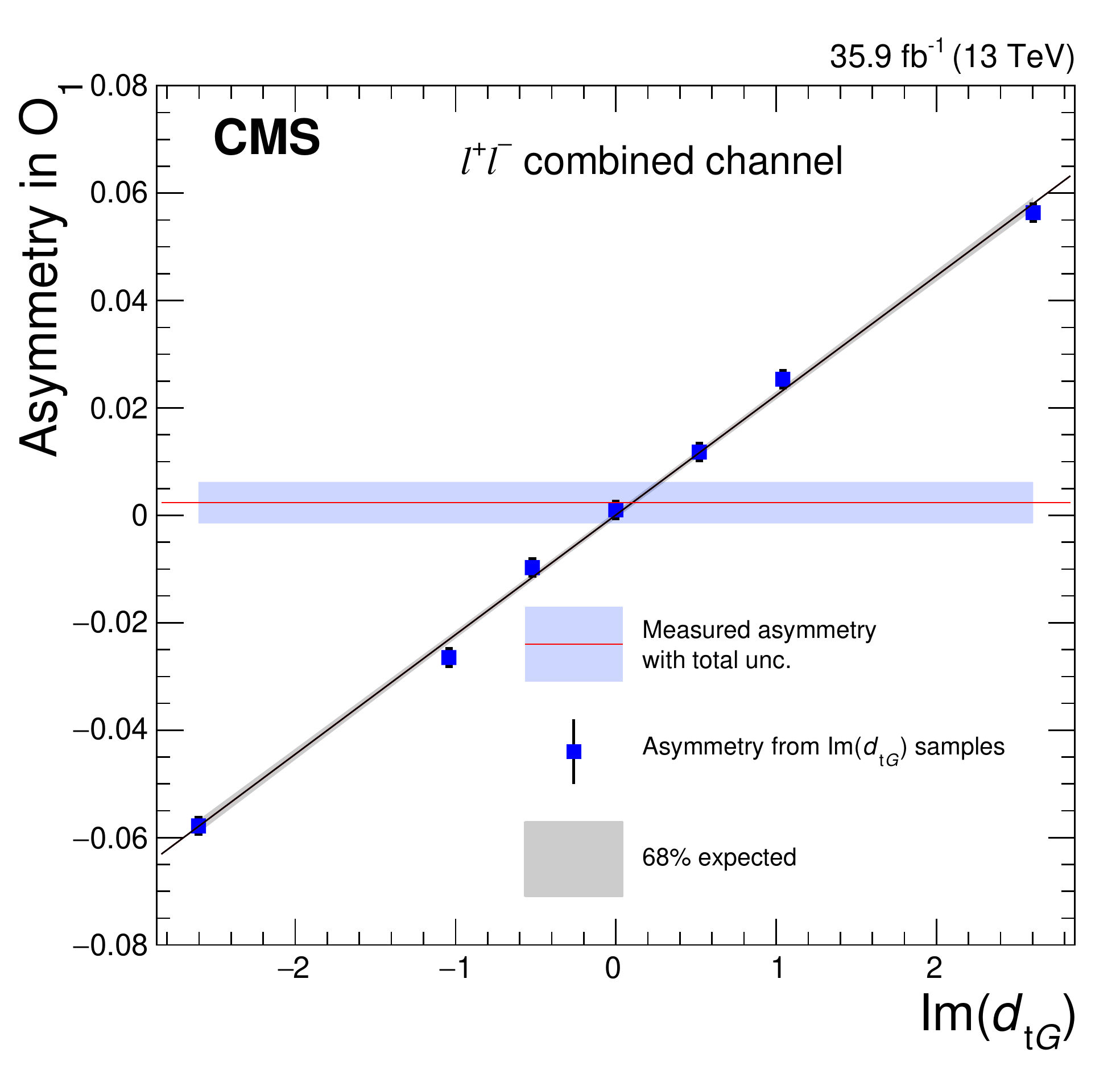}%
\hspace{0.05\textwidth}%
\includegraphics[width=0.45\textwidth]{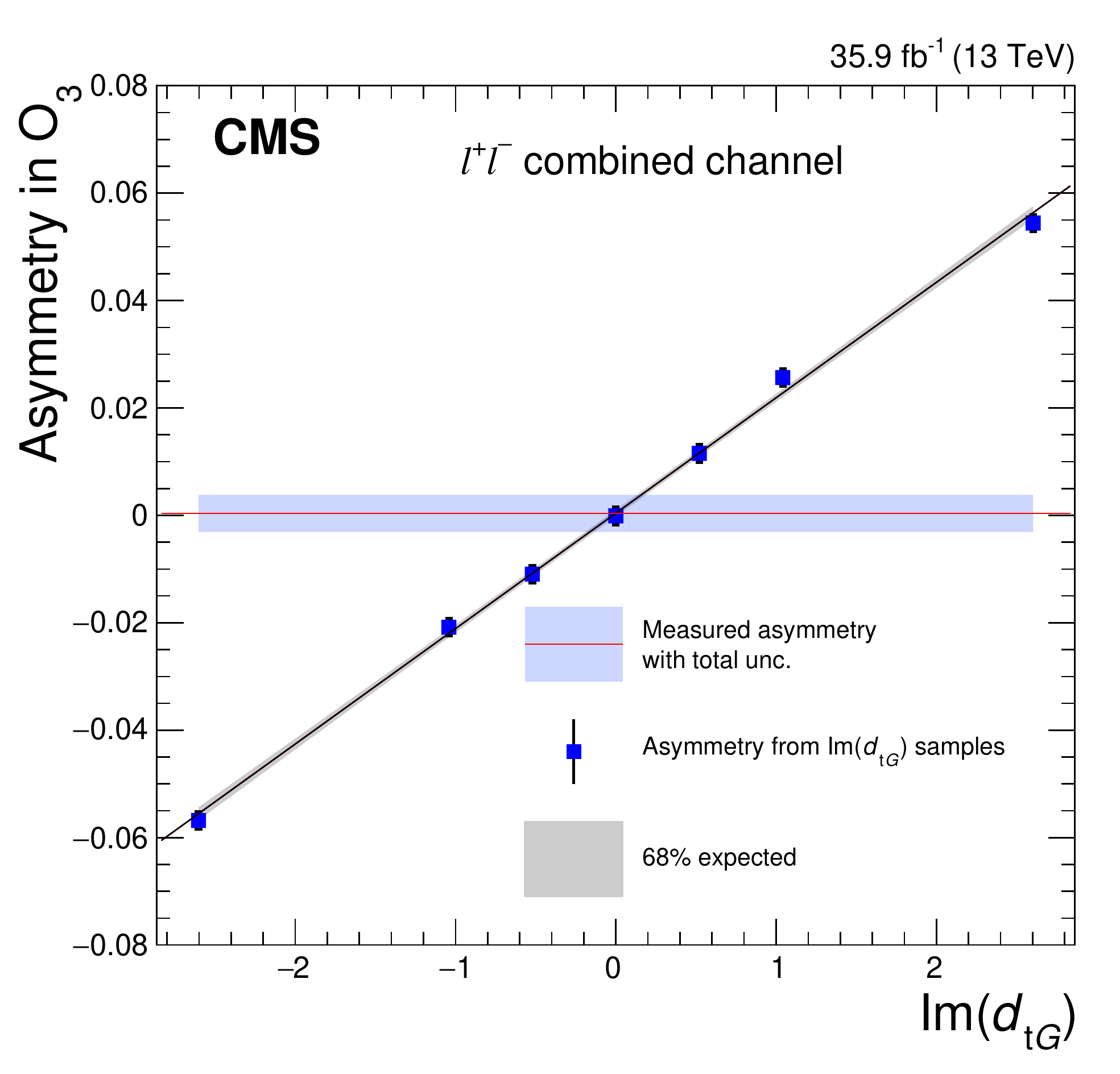}
\caption{Asymmetries as a function of \ImdtG for \ObsOne (left) and \ObsThree (right), for the combined dilepton channels.
The band corresponds to the uncertainties at the 68\% confidence level, of the linear fits.
The square markers are the asymmetries measured using simulated samples corresponding to the different \ImdtG values.
The horizontal line indicates the measured asymmetry, and the shaded region reflects the total statistical and systematic uncertainty.
}
\label{fig:cedm}
\end{figure}

Figure~\ref{fig:cedm} shows the expected asymmetries as a function of \ImdtG.
Within one standard deviation, the measured asymmetries are linearly proportional to the input CEDM values for both \ObsOne and \ObsThree. The relation
between the asymmetry $A$ and the CEDM can be written as
\begin{linenomath}\begin{equation}
\label{eq:asy_dtg}
A = a\, \ImdtG + b.
\end{equation}\end{linenomath}
The parameters $a$ and $b$ are obtained from a least-squares fit to the values obtained from the simulated samples.
From this relation, \ImdtG is extracted from the measured asymmetry as
\begin{linenomath}\begin{equation}
\label{eq:dtg_asym}
\ImdtG =  \frac{A_{\text{measured}} - b}{a}.
\end{equation}\end{linenomath}
The uncertainty in the measured dimensionless CEDM \ImdtG is calculated using the full covariance matrix from the fit as
\begin{linenomath}\begin{equation}
\label{eq:uncer_dtg}
\Delta_{\ImdtG}^2=
\begin{pmatrix}
\dfrac{\partial \ImdtG}{\partial A} & \dfrac{\partial \ImdtG}{\partial b} & \dfrac{\partial \ImdtG}{\partial a}
\end{pmatrix}
\renewcommand{\arraystretch}{1.5}
\begin{pmatrix}
{\Delta}^2_{A} & 0 & 0 \\
0 & \Delta^2_{b} & \text{cov}(b,a) \\
0 & \text{cov}(a,b) & \Delta^2_{a}
\end{pmatrix}
\renewcommand{\arraystretch}{2.3}
\begin{pmatrix}
\dfrac{\partial \ImdtG}{\partial A} \\
\dfrac{\partial \ImdtG}{\partial b} \\
\dfrac{\partial \ImdtG}{\partial a}
\end{pmatrix},
\end{equation}\end{linenomath}
with
$\Delta_{\ImdtG}$ being the uncertainty in \ImdtG, and $\Delta_A$, $\Delta_a$, and $\Delta_b$ the uncertainties in $A$, $a$, and $b$.
The measured asymmetries for the combined dilepton channel are used to extract the CEDM for \ObsOne and \ObsThree.
The measured values for the dimensionless CEDM \ImdtG with their statistical and systematic uncertainties are given in Table~\ref{tab:comb_final_results}.

\begin{table}[!h]
\renewcommand{\arraystretch}{1.2}
\centering
\topcaption{The measured asymmetries of \ObsOne and \ObsThree, and dimensionless CEDM \ImdtG, extracted using the asymmetries in \ObsOne and \ObsThree, with their uncertainties.}
\begin{tabular}{ccc}
Observable & Asymmetry (${\times}10^{-3}$)& \ImdtG \\
\hline
\AsymOne & $2.4 \pm 2.8\stat \pm 2.8\syst $ & $0.10 \pm 0.12\stat \pm 0.12\syst$ \\
\AsymThree & $0.4 \pm 2.8\stat \pm 2.2\syst$ & $0.00 \pm 0.13\stat \pm 0.10\syst$ \\
\end{tabular}
\label{tab:comb_final_results}
\end{table}

The measured asymmetries and CEDMs are consistent with the expectation from the SM of a negligible CP asymmetry.
The measurement of the \ttbar spin correlation in the CMS experiment~\cite{PhysRevD.100.072002} also provided a CEDM value using \hatdt proposed in Ref.~\cite{Bernreuther:2015yna}. The value \hatdt from the measured \ttbar spin correlation is $-0.020<\hatdt<0.012$ at the 95\% confidence level.
The parameter \hatdt is related to \ImdtG as
\begin{linenomath}\begin{equation}
\label{eq:dtg_dhat}
\frac{\hatdt}{\mtop} = \frac{\sqrt{2}v}{\Lambda^2} \ImdtG.
\end{equation}\end{linenomath}
At a new physics scale $\Lambda = 1\TeV$, using Eqs.~\eqref{eq:cedm_L} and~\eqref{eq:dtg_dhat}, the measured \ImdtG values can be converted into \hatdt, resulting in $-0.014<\hatdt<0.027$ and $-0.019<\hatdt<0.019$ at the 95\% confidence level, extracted from the asymmetries of \ObsOne and \ObsThree, respectively.
Hence, the sensitivity of this analysis to the CEDM \hatdt value is similar to that of the \ttbar spin correlation analysis.

\section{Summary}
\label{sec:Summary}
Violations of CP symmetry are studied in top quark pair production in the dilepton final state. The analysis is based on proton-proton collisions at a center-of-mass energy of 13\TeV, collected by the CMS experiment and corresponding to an integrated luminosity of 35.9\fbinv.
The analysis uses two observables, \ObsOne and \ObsThree, which are related to the Levi-Civita tensor contracted with the four-momenta of the leptons,  the jets originating from the \PQb quarks, and the top quarks.
Asymmetries are measured in these observables and converted to measurements of the chromoelectric dipole moment (CEDM) of the top quark, represented by the dimensionless CEDM \ImdtG.
The measured \ImdtG based on the asymmetries of the \ObsOne and \ObsThree observables in the combined dilepton channels are
$0.10 \pm 0.12\stat \pm 0.12\syst$,
and
$0.00 \pm 0.13\stat \pm 0.10\syst$,
respectively.
These results are consistent with the expectation of the standard model (SM), since in the SM prediction, the size of CP violation and the CEDM (\ImdtG) is negligible. In this analysis, the extracted CEDMs are compared with the CEDM measured in the \ttbar spin correlation analysis~\cite{PhysRevD.100.072002}, and the sensitivity is found to be similar.

\begin{acknowledgments}

We congratulate our colleagues in the CERN accelerator departments for the excellent performance of the LHC and thank the technical and administrative staffs at CERN and at other CMS institutes for their contributions to the success of the CMS effort. In addition, we gratefully acknowledge the computing centers and personnel of the Worldwide LHC Computing Grid and other centers for delivering so effectively the computing infrastructure essential to our analyses. Finally, we acknowledge the enduring support for the construction and operation of the LHC, the CMS detector, and the supporting computing infrastructure provided by the following funding agencies: BMBWF and FWF (Austria); FNRS and FWO (Belgium); CNPq, CAPES, FAPERJ, FAPERGS, and FAPESP (Brazil); MES and BNSF (Bulgaria); CERN; CAS, MoST, and NSFC (China); MINCIENCIAS (Colombia); MSES and CSF (Croatia); RIF (Cyprus); SENESCYT (Ecuador); MoER, ERC PUT and ERDF (Estonia); Academy of Finland, MEC, and HIP (Finland); CEA and CNRS/IN2P3 (France); BMBF, DFG, and HGF (Germany); GSRI (Greece); NKFIH (Hungary); DAE and DST (India); IPM (Iran); SFI (Ireland); INFN (Italy); MSIP and NRF (Republic of Korea); MES (Latvia); LAS (Lithuania); MOE and UM (Malaysia); BUAP, CINVESTAV, CONACYT, LNS, SEP, and UASLP-FAI (Mexico); MOS (Montenegro); MBIE (New Zealand); PAEC (Pakistan); MES and NSC (Poland); FCT (Portugal); MESTD (Serbia); MCIN/AEI and PCTI (Spain); MOSTR (Sri Lanka); Swiss Funding Agencies (Switzerland); MST (Taipei); MHESI and NSTDA (Thailand); TUBITAK and TENMAK (Turkey); NASU (Ukraine); STFC (United Kingdom); DOE and NSF (USA).

\hyphenation{Rachada-pisek}
Individuals have received support from the Marie-Curie program and the European Research Council and Horizon 2020 Grant, contract Nos.\ 675440, 724704, 752730, 758316, 765710, 824093, 884104, and COST Action CA16108 (European Union); the Leventis Foundation; the Alfred P.\ Sloan Foundation; the Alexander von Humboldt Foundation; the Belgian Federal Science Policy Office; the Fonds pour la Formation \`a la Recherche dans l'Industrie et dans l'Agriculture (FRIA-Belgium); the Agentschap voor Innovatie door Wetenschap en Technologie (IWT-Belgium); the F.R.S.-FNRS and FWO (Belgium) under the ``Excellence of Science -- EOS" -- be.h project n.\ 30820817; the Beijing Municipal Science \& Technology Commission, No. Z191100007219010; the Ministry of Education, Youth and Sports (MEYS) of the Czech Republic; the Hellenic Foundation for Research and Innovation (HFRI), Project Number 2288 (Greece); the Deutsche Forschungsgemeinschaft (DFG), under Germany's Excellence Strategy -- EXC 2121 ``Quantum Universe" -- 390833306, and under project number 400140256 - GRK2497; the Hungarian Academy of Sciences, the New National Excellence Program - \'UNKP, the NKFIH research grants K 124845, K 124850, K 128713, K 128786, K 129058, K 131991, K 133046, K 138136, K 143460, K 143477, 2020-2.2.1-ED-2021-00181, and TKP2021-NKTA-64 (Hungary); the Council of Science and Industrial Research, India; the Latvian Council of Science; the Ministry of Education and Science, project no. 2022/WK/14, and the National Science Center, contracts Opus 2021/41/B/ST2/01369 and 2021/43/B/ST2/01552 (Poland); the Funda\c{c}\~ao para a Ci\^encia e a Tecnologia, grant CEECIND/01334/2018 (Portugal); the National Priorities Research Program by Qatar National Research Fund;  MCIN/AEI/10.13039/501100011033, ERDF ``a way of making Europe", and the Programa Estatal de Fomento de la Investigaci{\'o}n Cient{\'i}fica y T{\'e}cnica de Excelencia Mar\'{\i}a de Maeztu, grant MDM-2017-0765 and Programa Severo Ochoa del Principado de Asturias (Spain); the Chulalongkorn Academic into Its 2nd Century Project Advancement Project, and the National Science, Research and Innovation Fund via the Program Management Unit for Human Resources \& Institutional Development, Research and Innovation, grant B05F650021 (Thailand); the Kavli Foundation; the Nvidia Corporation; the SuperMicro Corporation; the Welch Foundation, contract C-1845; and the Weston Havens Foundation (USA).

\end{acknowledgments}

\bibliography{auto_generated}

\cleardoublepage \appendix\section{The CMS Collaboration \label{app:collab}}\begin{sloppypar}\hyphenpenalty=5000\widowpenalty=500\clubpenalty=5000\input{TOP-18-007-public-authorlist.tex}\end{sloppypar}
%%% END EDITABLE REGION %%%
% skeleton_end
\end{document}

%% file: TOP-18-007-public-authorlist.tex
\cmsinstitute{Yerevan Physics Institute, Yerevan, Armenia}
{\tolerance=6000
A.~Tumasyan\cmsAuthorMark{1}\cmsorcid{0009-0000-0684-6742}
\par}
\cmsinstitute{Institut f\"{u}r Hochenergiephysik, Vienna, Austria}
{\tolerance=6000
W.~Adam\cmsorcid{0000-0001-9099-4341}, T.~Bergauer\cmsorcid{0000-0002-5786-0293}, M.~Dragicevic\cmsorcid{0000-0003-1967-6783}, A.~Escalante~Del~Valle\cmsorcid{0000-0002-9702-6359}, R.~Fr\"{u}hwirth\cmsAuthorMark{2}\cmsorcid{0000-0002-0054-3369}, M.~Jeitler\cmsAuthorMark{2}\cmsorcid{0000-0002-5141-9560}, N.~Krammer\cmsorcid{0000-0002-0548-0985}, L.~Lechner\cmsorcid{0000-0002-3065-1141}, D.~Liko\cmsorcid{0000-0002-3380-473X}, I.~Mikulec\cmsorcid{0000-0003-0385-2746}, F.M.~Pitters, J.~Schieck\cmsAuthorMark{2}\cmsorcid{0000-0002-1058-8093}, R.~Sch\"{o}fbeck\cmsorcid{0000-0002-2332-8784}, M.~Spanring\cmsorcid{0000-0001-6328-7887}, S.~Templ\cmsorcid{0000-0003-3137-5692}, W.~Waltenberger\cmsorcid{0000-0002-6215-7228}, C.-E.~Wulz\cmsAuthorMark{2}\cmsorcid{0000-0001-9226-5812}, M.~Zarucki\cmsorcid{0000-0003-1510-5772}
% ...
\par}
\vskip\cmsinstskip
\dag:~Deceased\\
$^{1}$Also at Yerevan State University, Yerevan, Armenia\\
$^{2}$Also at TU Wien, Vienna, Austria\\
% ...
